\documentclass[aps,prd,twoside,titlepage,nofootinbib] {revtex4-2}

\usepackage{amssymb} 
\usepackage{graphicx}
\usepackage{grffile}
\usepackage{amsmath}
\usepackage[labelformat=simple]{subcaption}
\usepackage[table]{xcolor}
\usepackage{lineno}
\usepackage[breaklinks=true]{hyperref}
\usepackage{multirow}

\newcommand{\gevc}{\mbox{$\mathrm{GeV/c}$}}

\newcommand {\mee} {M_{ee}}

\newcommand {\normupsAll} {\frac{N_{\varUpsilon(1S+2S+3S)}}{\left\langle N_{\varUpsilon(1S+2S+3S)} \right\rangle}}

\newcommand {\meanmult} {\left\langle N_{ch} \right\rangle}
\newcommand {\meanUps} {\left\langle N_{\varUpsilon} \right\rangle}

\begin{document}


\title{Measurements of $\varUpsilon$ States Production in \textit{$p+p$} Collisions at $\sqrt{s} = 500\:\mathrm{GeV}$ with STAR: Cross Sections, Ratios, and Multiplicity Dependence}

\author{The STAR Collaboration}

\begin{abstract}
We report measurements of $\varUpsilon(1S)$, $\varUpsilon(2S)$ and $\varUpsilon(3S)$ production in \textit{$p+p$} collisions at $\sqrt{s}=500\:\mathrm{GeV}$ by the STAR experiment in year 2011, corresponding to an integrated luminosity $\mathcal{L}_{int}=13\:\mathrm{pb^{-1}}$.
The results provide precise cross sections, transverse momentum ($p_{T}$) and rapidity ($y$) spectra, as well as cross section ratios for $p_{\mathrm{T}}<10\:\gevc$ and $|y|<1$.
The dependence of the $\varUpsilon$ yield on charged particle multiplicity has also been measured, offering new insights into the mechanisms of quarkonium production. 
The data are compared to various theoretical models: the Color Evaporation Model (CEM) accurately describes the $\varUpsilon(1S)$ production, while the Color Glass Condensate + Non-relativistic Quantum Chromodynamics (CGC+NRQCD) model overestimates the data, particularly at low $p_{T}$.
Conversely, the Color Singlet Model (CSM) underestimates the rapidity dependence.
These discrepancies highlight the need for further development in understanding the production dynamics of heavy quarkonia in high-energy hadronic collisions.
The trend in the multiplicity dependence is consistent with CGC/Saturation and String Percolation models or $\varUpsilon$ production happening in multiple parton interactions modeled by PYTHIA8.
\end{abstract}

\preprint{Intended for submission to Physical Review D}


\maketitle

\section{Introduction}
\label{sec:int}

Heavy quarkonium states, such as $\varUpsilon$ mesons consisting of $b\bar{b}$ (bottom and anti-bottom quark), are vital tools for probing gluon content and dynamics in nucleons and nuclei~\cite{bib:Jpsi:TSSA,bib:STAR:Jpsi:dAu:gluonic}, and in a quark-gluon plasma (QGP)~\cite{bib:Ups:CMS:Seq, bib:Matsui1986}.
Despite extensive experimental and theoretical efforts, the mechanism of quarkonium production remains poorly understood.
The process involves a hard scattering event producing a heavy quark-antiquark pair, followed by non-perturbative evolution and hadronization into a bound state that can occur through either a color singlet (CS) or color octet (CO) intermediate state.
While the perturbative QCD (pQCD)~\cite{bib:pQCD:1, bib:pQCD:2} framework accurately describes the initial hard scattering. The subsequent bound state formation processes are modeled using different approaches, such as the Color Singlet Model (CSM)~\cite{bib:Onium:CS:1980, bib:Onium:CS:JpsiUps, bib:Onium:CS:Baier}, Non-Relativistic QCD (NRQCD)~\cite{bib:Onium:CO:QCDana,bib:Onium:CO:QCDanaErr} with both CS and CO states~\cite{bib:Onium:CO:Bmeson,bib:Onium:CO:QCDana,bib:Onium:CO:QCDanaErr}, and the Improved Color Evaporation Model (CEM)~\cite{bib:Jpsi:CEM:Fritzsch}.
Each model offers a different perspective, leading to ongoing debates about the correct production mechanism~\cite{bib:Kosarzewski:2022enn, bib:lansberg_newObs}.
Compared to $J/\psi$, $\varUpsilon$ states are a cleaner probe, because of less feed-down from excited states and offer an opportunity to study each state separately.

Precise $\varUpsilon$ production measurements have been conducted at the Tevatron~\cite{bib:UpsCDF, bib:UpsCDF_1960, bib:Ups:CDFratio} and the Large Hadron Collider (LHC)~\cite{bib:ALICE_states, bib:Ups:CMSspectra,bib:UpsCMS_2010,bib:Ups:CMS:diffXsec,bib:Ups:CMS:Xsec,bib:Ups:AtlasRatio,bib:Ups:LHCbRatio,bib:Ups:LHCb8Tev,bib:Ups:CMS:DPS, bib:Ups:CMS:DPS2, bib:Ups:CDFratio, bib:UpsCDF}, with additional data available from lower-energy experiments~\cite{bib:Ups:CFSpp, bib:UpsISR, bib:Ups:CCOR2, bib:Ups:PHENIX:ppAuAu, bib:Ups:STAR:pp, bib:Ups:STAR:dAu}, which offer lower precision due to low $\varUpsilon$ production rate. However, more data are needed at intermediate energies (between $19.4\:\mathrm{GeV}$ and $1\:\mathrm{TeV}$) to refine and constrain production models. 
The description of $\varUpsilon$ spectra depend on long-distance matrix elements (LDMEs), which vary across different studies of LHC and Tevatron data, indicating a need for more constraints.
Further insights could be gained by examining quarkonium production versus charged particle multiplicity, associated production, and polarization studies~\cite{bib:lansberg_newObs}.

Recent measurements of $J/\psi$~\cite{bib:ALICE:JpsiEventAct,bib:Jpsi:ALICE:EvAct13TeV,bib:Jpsi:pp:STAR:mult} by ALICE, STAR and $\varUpsilon$~\cite{bib:Ups:CMSactivity} production by CMS versus charged particle multiplicity in \textit{$p+p$} collisions have shown a strong increase of quarkonium yields in high-multiplicity events, which provide insights into the interplay of hard and soft QCD processes. Potential explanations include quenching interactions of overlapping strings in String Percolation~\cite{bib:PercolationJpsi} model, a 3-gluon contribution to $\varUpsilon$ production in Color Glass Condensate CGC/Saturation~\cite{bib:EvAct:CGCwatanabe,bib:EventAct:Jpsi:3pom,bib:EventAct:QQ:CGC}, or $\varUpsilon$ production in multiple parton interactions (MPI)~\cite{bib:ALICE:CBactivity}, and Coherent Particle Production~\cite{bib:Jpsi:EvAct:Kopeliovich:CPP,bib:Jpsi:EvAct:Kopeliovich}, though further studies are necessary.

In this paper, we present a comprehensive study of $\varUpsilon$ states production, including differential cross sections versus $p_{\mathrm{T}}$ and rapidity ($y$), cross section ratios, and the dependence on charged particle multiplicity, measured by the STAR experiment in \textit{$p+p$} collisions at $\sqrt{s} = 500\:\mathrm{GeV}$.
These results provide important constraints on the assumptions in various production model calculations and offer new insights into the mechanisms underlying heavy quarkonium production. 

This paper is organized as follows: Sections~\ref{sec:int} and ~\ref{sec:exp} provide an introduction and description of the experimental setup and data collection, respectively. Section~\ref{sec:ana} outlines the data analysis techniques. Section~\ref{sec:res} presents the results, compares them to theoretical model calculations, and discusses their implications. Finally, Section~\ref{sec:summ} provides a summary of our findings.

\section{Experimental setup and data taking}
\label{sec:exp}

The data used in this study were recorded by the STAR experiment during the 2011 data-taking run ~\cite{bib:tech_STAR} and consists of \textit{$p+p$} events at $\sqrt{s}=500\:\mathrm{GeV}$.
These events were triggered by requiring a minimum energy $E\gtrapprox4.6\:\mathrm{GeV}$ deposited in a single channel in the Barrel Electromagnetic Calorimeter (BEMC)~\cite{bib:tech_BEMC}, referred to as an L0 triggering tower, and a coincidence of hits in both the east and west Beam-Beam Counters (BBCs)~\cite{bib:tech:STAR:BBCpol}.
Such conditions select a valid \textit{$p+p$} collision and trigger on possible $\varUpsilon$ candidate events decaying via the dielectron channel. The BEMC has full $2\pi$ azimuthal acceptance within pseudo-rapidity~\footnote{$\eta = -\mathrm{ln}(\mathrm{tan}(\theta/2))$} $|\eta|<1$, while the BBCs provide coverage in $3.4<|\eta|<5.0$. Tracking and particle identification using energy loss information $dE/dx$ are performed using the information from the Time Projection Chamber (TPC)~\cite{bib:tech_STARTPC} within $|\eta|<1$ (with reduced efficiency up to $|\eta|<1.8$) and $2\pi$ coverage in azimuthal angle. The Time of Flight detector (TOF)~\cite{bib:tech:STAR:TOFp} is used in this study to remove pile-up tracks for the measurement of charged particle multiplicity. The Shower Maximum Detector~\cite{bib:tech_BEMC} (SMD) is a part of the BEMC and is positioned at the expected location of the maximum of the electromagnetic shower within the calorimeter tower. The track projections to the SMD layer are used to match TPC tracks with the calorimeter energy deposits.

\section{Data analysis}
\label{sec:ana}

In order to measure the production cross sections, the $\varUpsilon$ signal has to be reconstructed and corrected for the reconstruction efficiency. The charged particle multiplicity dependence, is studied by correcting the measured multiplicity distributions for $\varUpsilon$ and minimum-bias events using tracking efficiency studied with a Monte-Carlo (MC) simulation and an unfolding procedure~\cite{bib:RooUnfoldMan}.

Analysis of the data starts by selecting events with a reconstructed primary collision vertex position along the beam axis with $|V_{z}|<40\:\mathrm{cm}$. This is to ensure uniform acceptance by requiring collisions happen in the center of the TPC fiducial volume.
The number of events accepted for analysis after the primary vertex selection is $92\:\mathrm{M}$, which corresponds to an integrated luminosity of $13\:\mathrm{pb^{-1}}$.

Reconstruction of the $\varUpsilon(nS)$ states is done via the dielectron channel, which has a branching ratio of $B^{\varUpsilon(1S)}_{ee} = 2.39\pm0.08\%$, $B^{\varUpsilon(2S)}_{ee} = 1.91\pm0.16\%$ and $B^{\varUpsilon(3S)}_{ee} = 2.18\pm0.20\%$~\cite{PDG:2024} for $\varUpsilon(1S)$, $\varUpsilon(2S)$ and $\varUpsilon(3S)$ states, respectively.
Electron tracks are measured by the TPC and must satisfy $p_{\mathrm{T}}>0.2\:\mathrm{GeV/c}$ with at least $20$ measured space points (out of a possible maximum of 45 at mid-rapidity) are required in the track fitting to ensure acceptable momentum resolution.
Uniform track reconstruction efficiency across the TPC acceptance is ensured by selecting tracks with $|\eta|<1.0$.
In addition, tracks originating from the primary vertex are selected by calculating the distance of closest approach ($DCA$) between the primary vertex and the track trajectory and choosing tracks with $DCA<3\:\mathrm{cm}$.
Furthermore, split tracks are removed by requiring the ratio of track fit points to the maximum possible number at a given $\eta$ and $p_{\mathrm{T}}$ to be greater than $0.52$. This makes sure it is greater than half, which would be less than half in case the tracking algorithm reconstructed a single particle as two tracks.

\subsection{Event charged particle multiplicity measurement}

For the purpose of investigating the dependence of $\varUpsilon$ yields on charged particle multiplicity ($N_{ch}$), a stable measure of the number of tracks is required.
An acceptable measure of track multiplicity must be unaffected by pile-up tracks, which may artificially increase $N_{ch}$, and be insensitive to event-by-event fluctuations.
The latter can be satisfied by accepting tracks with only basic quality selection. Thus, similar criteria to those outlined above for electron tracks are used, except for the few relaxed ones, which are: number of fit points, fit points to maximum ratio and $DCA$. The number of required track fit points is reduced to $\geq 15$, which reduces track number fluctuations. On the other hand, large pile-up requires a stricter selection of $DCA<0.5\:\mathrm{cm}$ and for the track to have produced a hit in the TOF. The TOF provides accurate timing information and can be used to reject particle tracks coming from different bunch crossings. Charged particle multiplicity measured using tracks with TOF information is referred to as ``TOF multiplicity" in this study. 
The electrons and positrons coming from $\varUpsilon$ decays are included in the $N_{ch}$ calculation.
The minimum-bias $N_{ch}$ distribution is obtained with the same criteia listed above from a separate low-luminosity dataset which has lower pile-up contamination and enables a wider $DCA<1.0\:\mathrm{cm}$ requirement. 

\subsection{Electron identification and selection}

The electron candidates are selected using $dE/dx$ information measured by the TPC by calculating an $n\sigma_{e}$ variable according to Eq.~\ref{nsigmaeq}, which quantifies the deviation of measured $dE/dx_{\mathrm{meas}}$ from the expected value, $dE/dx_{\mathrm{exp}}$ according to the Bichsel function~\cite{bib:STAR_Bichsel}, scaled by the $dE/dx$ resolution $\sigma_{dE/dx}$.

\begin{equation}
\label{nsigmaeq}
n\sigma_{e} = \frac{\mathrm{ln}(\frac{dE/dx_{\mathrm{meas}}}{dE/dx_{\mathrm{exp}}})}{\sigma_{dE/dx}}
\end{equation}

Electron candidates are selected with $-1.2<n\sigma_{e}<3.0$, which has a high efficiency for electrons while minimizing the contamination from pions.
The BEMC is also used to identify electrons using the $E_{clu}/p$ ratio and shower shape, where $E_{clu}$ is the energy of the BEMC cluster matched to the TPC track, and $p$ is the momentum of the track as measured in the TPC. Matching to the BEMC clusters is done by projecting TPC tracks to towers (each with dimensions of $0.05\times0.05$ in $\Delta\eta$ and $\Delta\phi$) in the BEMC and forming clusters by adding two adjacent towers with the highest energy around each projection (eg. max 3 towers in each cluster).
The distance between the energy-weighted center of a cluster and the track projection to the SMD layer, $R_{SMD}=\sqrt{(\Delta\eta)^{2}+(\Delta\phi)^{2}}$, is calculated and used to determine if the TPC track projection matches the BEMC cluster by imposing a requirement of $R_{SMD}<0.028$.
The SMD layer is located at a distance of $231.723\:\mathrm{cm}$ from the beam.
Electrons are expected to deposit most of their energy in the BEMC, in contrast to hadrons. This motivates a selection of ratio of cluster energy to track momentum of $0.55<\frac{E_{clu}}{p}<1.45$.
In order to further reject hadrons, information on shower characteristics is used.
Electrons should deposit a large fraction of their energy in a single tower, so a ratio of energy contained in the matched tower $E_{tow}$ to the energy of the entire cluster $\frac{E_{tow}}{E_{clu}}>0.5$ is used to select such clusters.

\subsection{Upsilon signal reconstruction}

\begin{figure}[h!]
\begin{center}
	\includegraphics[width=0.5\textwidth]{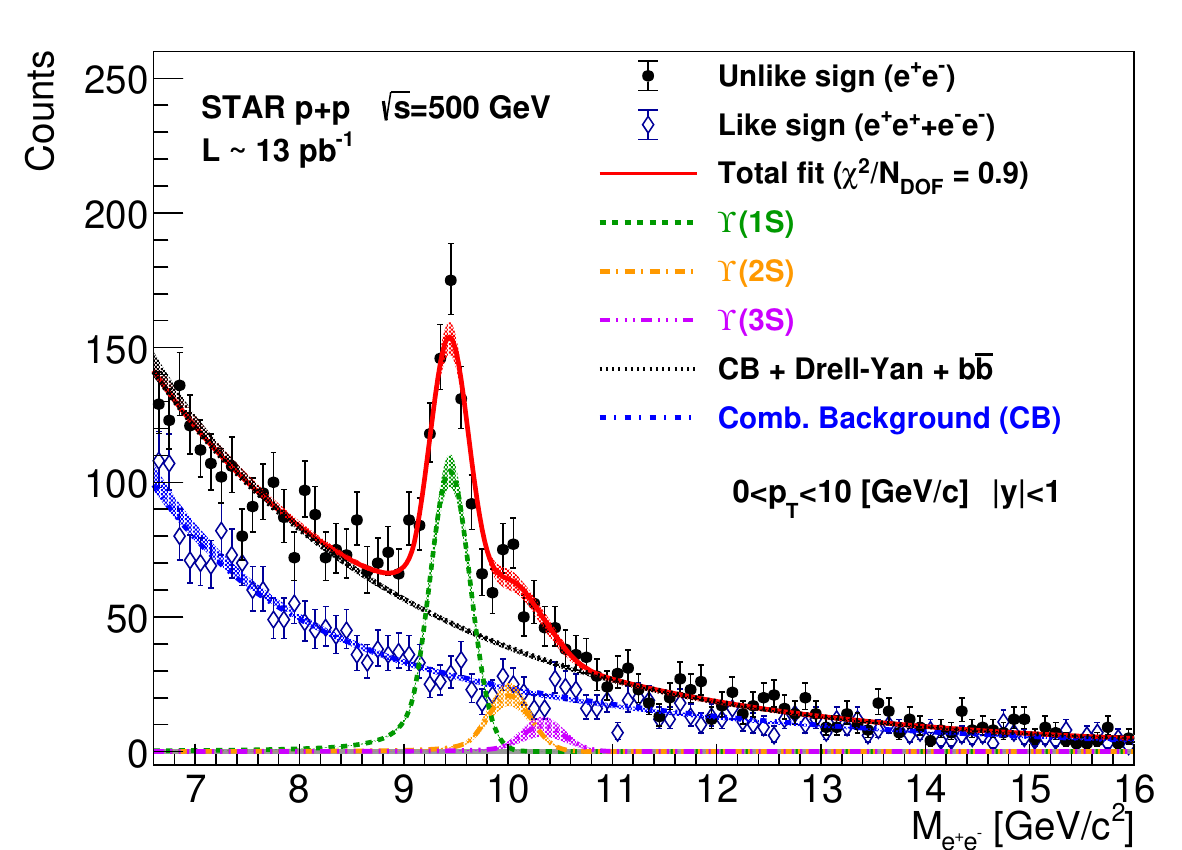}\\
\end{center}
\caption[Upsilon signal fit] 
{ \label{Fig:Fit}  
	Invariant mass $\mee$ distribution for unlike-sign (black full circles) and like-sign (blue hollow diamonds) electron pairs.
	The curves correspond to combinatorial background (blue dashed-dotted line), correlated background (black dotted line),
	$\varUpsilon(1S)$ (green), $\varUpsilon(2S)$ (orange), and $\varUpsilon(3S)$ (purple).
	The total (red) is a sum of the above components. Each curve has a corresponding uncertainty band obtained via. MC sampling technique and includes correlations between parameters.
}
\end{figure}

In order to reconstruct the $\varUpsilon$ signal, electron and positron candidates are made into pairs to reconstruct raw Upsilon candidates. One of the electron~\footnote{electrons means both electron and positron in the text unless specified} tracks has to be matched to the high energy BEMC tower (so-called ``High Tower") which fired the L0 trigger. A $p_{T}>1\:\gevc$ cut is required for the second electron partner, which reduces combinatorial background.
Electrons and positrons that pass all the selection requirements are paired and their invariant mass is calculated. 
In addition to the $\varUpsilon$ candidate $e^{+}e^{-}$ pairs, there are also contributions from combinatorial and correlated residual backgrounds.
Combinatorial background arises from randomly paired electrons or positrons and is estimated using a sum of like sign pairs ($e^{+}e^{+}$ or $e^{-}e^{-}$). The residual background consists of the $b\bar{b}$ continuum and Drell-Yan processes, but the latter contribution is small~\cite{bib:vogt:private}.
Fig.~\ref{Fig:Fit} shows the invariant mass of unlike-sign (black full circles) and like-sign (blue hollow diamonds) pairs. In order to extract the signal, these distributions in $6.6<\mee<16\:\mathrm{GeV/c^{2}}$ are fitted simultaneously with an unbinned likelihood method using RooFit~\cite{bib:RooFitMan}. The combinatorial background is modeled by an exponential function $f_{CB}=N_{CB}\cdot exp(\frac{-\mee}{T_{CB}})$ (blue curve) while the correlated background is described by a power-law function $f_{b\bar{b}+DY}=N_{b\bar{b}+DY}\frac{\mee^{A}}{(1+\frac{\mee}{B})^{C}}$ (black curve). The correlated background shape is determined using a PYTHIA8~\cite{bib:tools:PYTHIA8.2} simulation and serves as a constraint on the fit parameters. The shapes of each of the $\varUpsilon(nS)$ states are modeled with Crystal Ball functions~\cite{bib:CBfunction}, whose parameters (position, width, tail) are fixed using a full STAR detector simulation with $\varUpsilon(nS) \rightarrow e^{+}e^{-}$ decays embedded into raw data and reconstructed using the same selection criteria as outlined above.
Such approach allows for the correct modeling of signal widths, which is caused by the TPC momentum resolution.
The total fit is a sum of all these components (red curve).
Finally, the raw signal of each of the $\varUpsilon(nS)$ states is obtained directly from the fits.
The yields obtained in this way are later compared to those calculated with the bin counting method and the difference is taken as a systematic uncertainty.
The bin counting method integrates the histogram directly and relies on the fits for background subtraction.
It should be noted that the initial parameters for ratios of each $\varUpsilon$ state are set to the global fit values~\cite{bib:Ups:Ratios} and allowed to vary.

\subsection{Efficiency corrections}
\label{EffCorr}

\begin{figure}[h!]
\begin{center}
	\begin{subfigure}{0.49\textwidth}
	\includegraphics[width=1.0\textwidth]{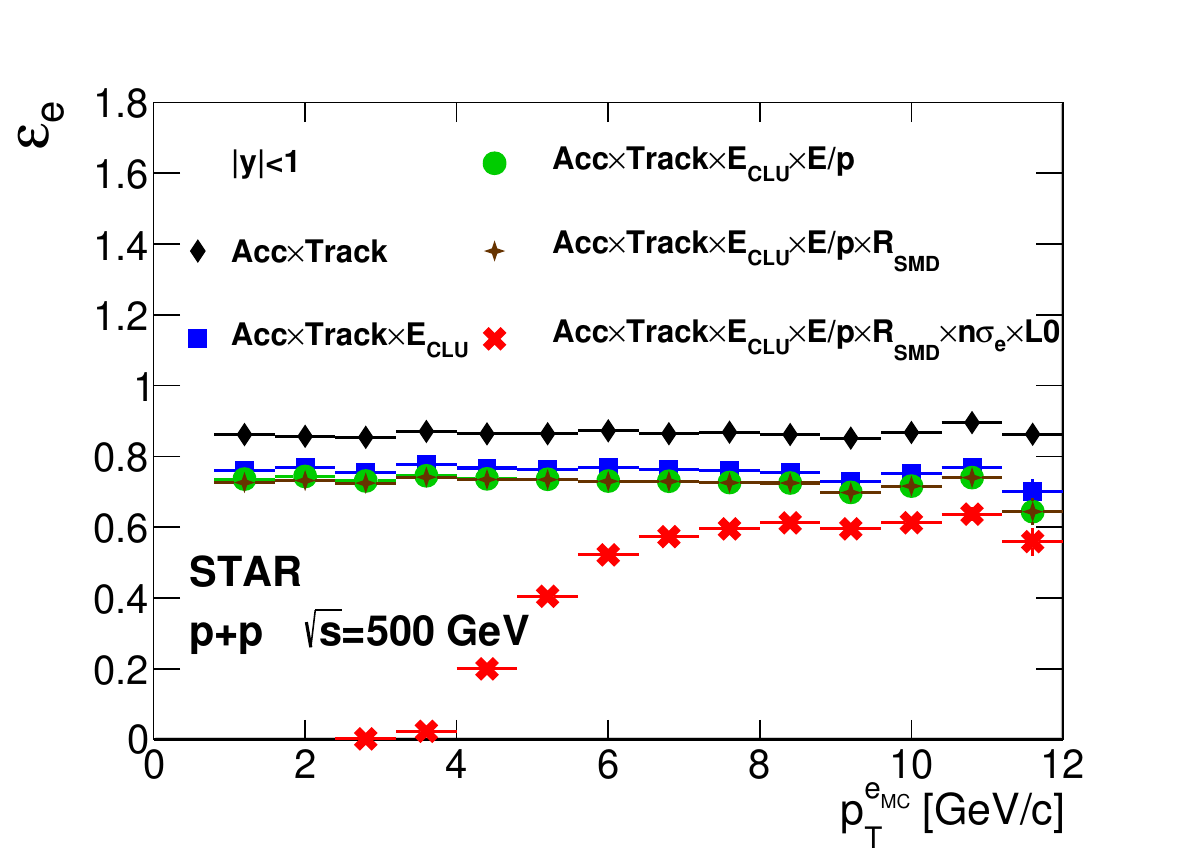}
		\caption[]{} \label{Fig:Eff:Ele}
	\end{subfigure}	
	\begin{subfigure}{0.49\textwidth}
	\includegraphics[width=1.0\textwidth]{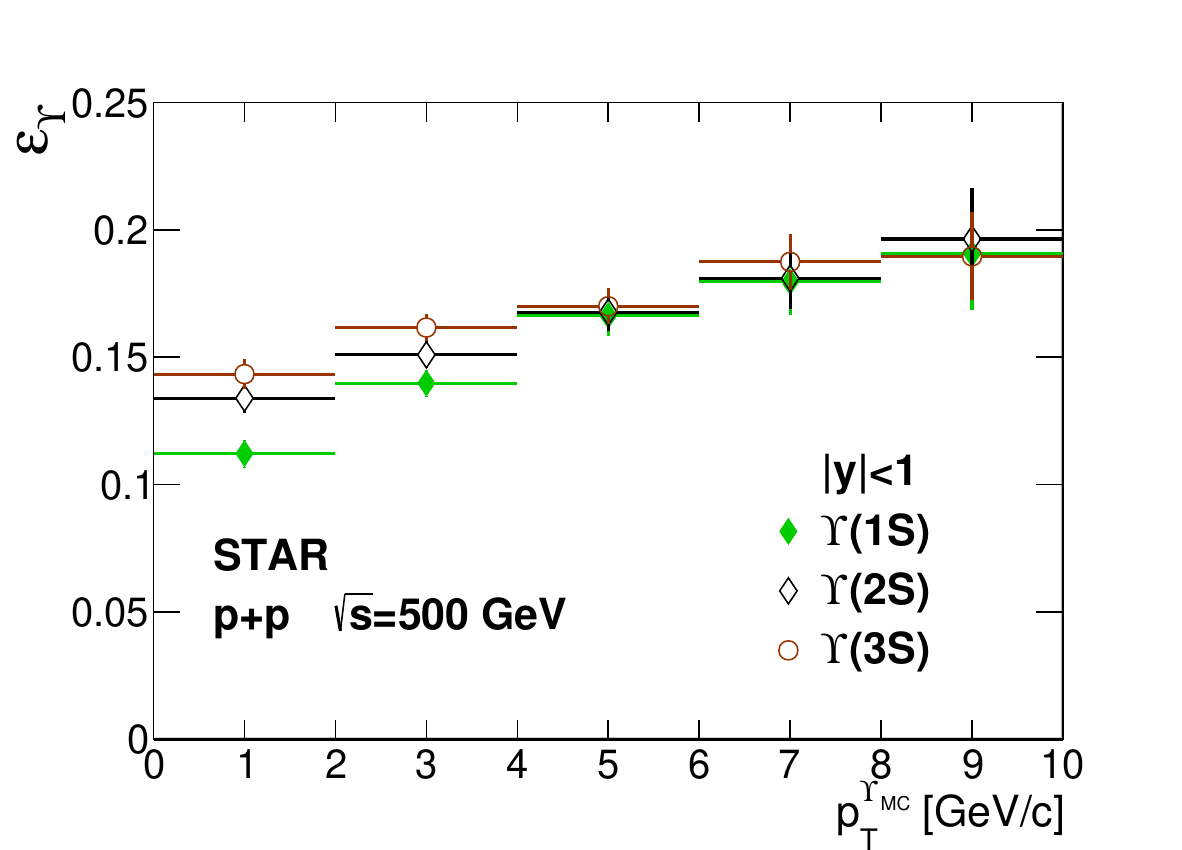}
		\caption[]{} \label{Fig:Eff:Ups}
	\end{subfigure}
\end{center}
\caption[Reconstruction efficiencies] 
{ \label{Fig:Eff}  
(a) Electron efficiencies vs. $p_{\mathrm{T}}^{e_{MC}}$. Illustrated is the combined effect of successive application of the acceptance and tracking efficiency (black diamonds), $E_{tow}/E_{clu}$ (blue rectangles), $E_{clu}/p$ (green circles), $R_{SMD}$ (brown stars) and $n\sigma_{e}$ along with L0 High Tower match (red crosses) requirements.
(b) Reconstruction efficiencies of $\varUpsilon(1S)$ (closed diamonds), $\varUpsilon(2S)$ (open diamonds), and $\varUpsilon(3S)$ (open circles) vs. $p_{\mathrm{T}}^{\varUpsilon_{MC}}$.
}
\end{figure}

In order to obtain the corrected $\varUpsilon(nS)$ signal, the $\varUpsilon(nS)$ reconstruction efficiency has to be studied. Three separate samples of simulated $\varUpsilon(nS) \rightarrow e^{+}e^{-}$ decays are generated ($\varUpsilon(1S)$, $\varUpsilon(2S)$, $\varUpsilon(3S)$), embedded into raw data, and reconstructed using a full simulation of the STAR detector geometry and response.
This accounts for different running conditions during the data taking period and allow for a controlled estimate of reconstruction efficiency for a known number of initial $\varUpsilon(nS) \rightarrow e^{+}e^{-}$ decays. The efficiency is calculated for single electrons by comparing the number of input electrons to those reconstructed from the simulated output.
The simulated $\varUpsilon(nS)$ are assumed to be unpolarized~\cite{bib:Ups:CDF:pol}, which effect is later studied as part of systematic uncertainties.
Fig.~\ref{Fig:Eff:Ele} shows the combined effects of acceptance and tracking efficiency (black diamonds), the $E_{tow}/E_{clu}$ (blue rectangles), the $E_{clu}/p$ (green circles), the $R_{SMD}$ (brown stars), and finally the $n\sigma_{e}$ and the L0 High Tower match (red crosses) vs. MC electron transverse momentum $p_{\mathrm{T}}^{e_{MC}}$ efficiencies. The $n\sigma_{e}$ selection efficiency is estimated using a pure electron sample of $\gamma \rightarrow e^{+}e^{-}$ conversions in the real data and checking the fraction passing the $n\sigma_{e}$ requirement.
The $\varUpsilon(nS)$ reconstruction efficiency shown in Fig.~\ref{Fig:Eff:Ups} is obtained by folding in the single electron efficiencies, however the L0 High Tower requirement is only applied to the highest energy electron.

The measured charged particle multiplicity estimated using tracks matched to the TOF is also corrected for the track reconstruction efficiency to obtain the true charged particle multiplicity.
This correction is done by performing four iterations of a Bayesian unfolding procedure with the RooUnfold package~\cite{bib:RooUnfoldMan}, similar to the technique used by STAR to study the dependence of $J/\psi$ production on multiplicity~\cite{bib:Jpsi:pp:STAR:mult}. The response matrix, which relates the measured TOF multiplicity to the true $N_{ch}$ is estimated using PYTHIA8 simulations separately for $\varUpsilon$ and minimum-bias events. Additionally, a BBC trigger efficiency correction is applied. This efficiency is estimated by embedding simulated PYTHIA8 $\varUpsilon$ and minimum-bias events into real 2011 zero-bias and a separate simulation based on low-luminosity minimum-bias data, respectively. The low-luminosity minimum-bias data also provide a measured TOF multiplicity distribution for minimum-bias events.

\subsection{Systematic uncertainties}

\begin{table}[h]
\centering
\begin{tabular}{|m{3.2cm}cccccc|cc|}
\hline
\multicolumn{7}{|c|}{$p_{\mathrm{T}}$ $[\gevc]$} & \multicolumn{2}{c|}{$y$ $[1]$} \\
\hline
Uncertainty [$\pm\%$] & $0-10$ & $0-2$ & $2-4$ & $4-6$ & $6-8$ & $8-10$ & $|y|<0.5$ & $0.5<|y|<1.0$ \\
\hline
Raw yield extraction & $1.2$ & $4.1$ & $2.0$ & $1.0$ & $0.1$ & $5.9$ & $0.3$ & $2.8$ \\
\hline
Fixed $\frac{\varUpsilon(2S)}{\varUpsilon(3S)}$ & $0.3$ & $0.3$ & $0.9$ & $0.3$ & $2.8$ & $0.9$ & $0.7$ & $1.3$ \\
\hline
$p_{\mathrm{T}}$ smearing & $1.2$ & $1.2$ & $0.8$ & $0.5$ & $0.7$ & $0.5$ & $0.7$ & $0.9$ \\
\hline
Tracking efficiency vs. $p_{\mathrm{T}}$ & $1.3$ & $1.4$ & $1.3$ & $1.3$ & $1.4$ & $1.3$ & $1.3$ & $1.3$ \\
\hline
Polarization & $0.3$ & $1.5$ & $0.2$ & $0.4$ & $0.5$ & $0.5$ & $0.1$ & $1.2$ \\
\hline
Trigger & $7.6$ & $17.0$ & $8.0$ & $2.9$ & $1.4$ & $0.9$ & $9.4$ & $5.5$ \\
\hline
Total & $7.9$ & $17.6$ & $8.4$ & $3.4$ & $3.5$ & $6.2$ & $9.6$ & $6.0$ \\
\hline
	\end{tabular}
		\caption[Summary of systematic uncertainties on the $\varUpsilon(1S+2S+3S)$ cross section vs. $p_{\mathrm{T}}$ and $y$] 
		{ \label{Tab:Ana:Ups:Syst:SummPt}
		Summary of systematic uncertainties on the $\varUpsilon(1S+2S+3S)$ cross section as a functions of $p_{\mathrm{T}}$ and $y$.
		}
\end{table}

\begin{table}[h]
\centering
\begin{tabular}{|cc|}
\hline
Uncertainty & Effect [$\pm\%$] \\
\hline
Luminosity & $8$ \\
\hline
Vertex & $1$ \\
\hline
Tracking efficiency const. & $10$ \\
\hline
Acceptance & $3$ \\
\hline
$n\sigma_{e}$ & $3.6$ \\
\hline
	\end{tabular}
		\caption[Summary of global systematic uncertainties on the $\varUpsilon(1S+2S+3S)$ cross section] 
		{ \label{Tab:Ana:Ups:Syst:Global}
		Summary of $p_{T}$-correlated systematic uncertainties on the $\varUpsilon(1S+2S+3S)$ cross section.
		}
\end{table}

\begin{table}[h]
\centering
\begin{tabular}{|m{3cm}cccc|}
\hline
\multirow{2}{*}{Uncertainty [$\pm\%$]} & \multicolumn{4}{c|}{$\meanmult$} \\
 & $0-1$ & $1-2$ & $2-3$ & $3-8$ \\
\hline
Number of iterations & $ 1.3$ & $ 1.3$ & $ 0.5$ & $ 0.5$ \\
\hline
Reconstruction efficiency & $ 0.2$ & $ 0.6$ & $ 0.4$ & $ 3.8$ \\
\hline
Tracking efficiency const. & $ 9.2$ & $ 4.9$ & $ 2.8$ & $ 0.4$ \\
\hline
Tracking efficiency vs. $p_{\mathrm{T}}$  & $ 1$ & $ 1$ & $ 1$ & $ 14$ \\
\hline
$N_{ch}$ from NBD & $ 0.3$ & $ 1.8$ & $ 10$ & $ 6.6$ \\
\hline
Raw yield extraction & $ 1.1$ & $ 2.5$ & $ 0.2$ & $ 15.7$ \\
\hline
$p_{\mathrm{T}}$ smearing & $ 0.3$ & $ 0.3$ & $ 0.2$ & $ 1.8$ \\
\hline
Fixed $\frac{\varUpsilon(2S)}{\varUpsilon(3S)}$ & $ 0.4$ & $ 0.4$ & $ 2.7$ & $ 10.8$ \\
\hline
4Cx tune & $ 4$ & $ 0.2$ & $ 3.5$ & $ 13.1$ \\
\hline
Total & $ 10.3$ & $ 6.1$ & $ 11.4$ & $ 28.1$ \\
\hline
	\end{tabular}
		\caption[Summary of systematic uncertainties on the $\normupsAll$] 
		{ \label{Tab:Ana:Ups:Syst:Activity}
		Summary of systematic uncertainties on the $\normupsAll$.
		}
\end{table}

\begin{table}[h]
\centering
\begin{tabular}{|m{3.2cm}cccc|}
\hline
\multirow{2}{*}{Uncertainty [$\pm\%$]} & \multicolumn{4}{c|}{$\meanmult$} \\
 & $0-1$ & $1-2$ & $2-3$ & $3-8$ \\
\hline
Iterations & $ 0.1$ & $ 0.4$ & $ 0.3$ & $ 0.2$ \\
\hline
Tracking efficiency & $ 3.5$ & $ 3.7$ & $ 4.0$ & $ 3.6$ \\
\hline
$N_{ch}$ from NBD & $ 0.1$ & $ 2.7$ & $ 2.1$ & $ 2.6$ \\
\hline
4Cx tune & $ 1.9$ & $ 0.0$ & $ 0.6$ & $ 0.3$ \\
\hline
Total & $ 4.0$ & $ 4.6$ & $ 4.6$ & $ 4.5$ \\
\hline
	\end{tabular}
		\caption[Summary of systematic uncertainties on the multiplicity] 
		{ \label{Tab:Ana:Ups:Syst:ActivityMult}
		Summary of systematic uncertainties on the multiplicity.
		}
\end{table}

This subsection describes the study of systematic uncertainties related to the methods and assumptions in this analysis.
Each of the contributions is symmetrized assuming the largest variation. First, the uncertainty associated with the signal extraction method is investigated by comparing the combined $\varUpsilon$ yield obtained using bin counting and the yield obtained directly from the fit. The effect is found to range from $\pm0.1\%$ to $\pm5.9\%$ depending on $p_{\mathrm{T}}$, and for the integrated signal is $\pm1.2\%$. Another contribution is studied by fixing the $\frac{\varUpsilon(2S)}{\varUpsilon(3S)}$ ratio to the value listed by PDG~\cite{bib:PDG} during the fitting and it is found to range from $\pm0.3$ to $\pm2.8\%$ vs. $p_{\mathrm{T}}$, for a total effect of $\pm0.3\%$ on the integrated yield.
When obtaining the shape of the $\varUpsilon$ signal in the simulation, the electron momentum resolution was smeared with a gaussian of width $a\times p_{T}$ , where $a$ was optimized to describe the $J/\psi$ signal width. By varying the amount of additional smearing within uncertainties, the impact of this correction may be estimated and it is found to affect the integrated yield by $\pm1.2\%$, while the $p_{\mathrm{T}}$ dependent effect varies between $\pm0.5\%$ and $\pm1.2\%$.
The impact of the tracking efficiency estimation is studied by varying the number of track fitting points in the simulation by $\pm2$. This is motivated by the fact that the TPC response simulation used in the embedding may not be perfect and could potentially miss hits.
The outcome of this effect is found to be $\pm1.3\%$ on the integrated signal with very little $p_{\mathrm{T}}$ dependence.
An additional constant $5\%$~\cite{bib:STAR:TrkEffSyst} uncertainty is estimated on the single track reconstruction efficiency, affecting the signal by $10\%$ globally. It is thus included as a correlated uncertainty.
To study the impact from the assumption of no $\varUpsilon$ polarization, described by $w(\theta) = 1+\lambda cos^{2}(\theta)$ in the simulations, the original value of the polarization parameter $\lambda=0$ is varied between $\pm0.1$, which corresponds to the bounds of CDF data~\cite{bib:Ups:CDF:pol}. This affects the $p_{\mathrm{T}}$-dependent yield by $\pm0.2\%$ to $\pm1.5\%$ resulting in a total effect of $\pm0.3\%$ on the integrated yield.
Additionally, the trigger response may be not well simulated. This was investigated by comparing the ADC\footnote{Analog to digital conversion} count distributions between data and simulation. Upon comparison, data and simulation were found to differ by $3\%$~\cite{bib:Ups:STAR:dAu}, which, when shifting the ADC value in the efficiency calculation affects the yield by $\pm7.6\%$.
All of these components contribute to the total uncorrelated systematic uncertainty of $\pm7.9\%$.

The $p_{T}$-correlated systematic uncertainties are estimated as follows.
First, the integrated luminosity estimation uncertainty was evaluated by studying the efficiency of the BBC detector and the effect was found to be $\pm8\%$~\cite{bib:STAR:Drun09, bib:STAR:highptRaa, bib:STAR:Long2Spin}. Next, the vertex finding efficiency uncertainty was studied by comparing the number of events with a reconstructed vertex to the total number~\cite{bib:Ups:STAR:dAu} of simulated events, the difference is found to be $\pm1\%$. Acceptance variations due to fluctuating BEMC tower availability during the run, which is later applied in the simulation, yields an additional $\pm3\%$~\cite{bib:Ups:STAR:dAu} effect. The uncertainty on the $n\sigma_{e}$ efficiency calculation is estimated by taking the $3\sigma$ upper and lower limits of the fit to $p$ dependence of the efficiency and the effect is $\pm3.6\%$.

The uncertainties connected with the $N_{ch}$-dependent studies are also investigated.
Some of the aforementioned uncertainties also affect $N_{ch}$ dependence. To check the influence of the unfolding method, the number of iterations was varied between $3$, $6$ and $8$. It was found that more iterations did not provide an improvement and the effect ranges from $\pm 0.4\%$ to $\pm 1.3\%$ on $\frac{N_{\varUpsilon}}{\langle N_{\varUpsilon} \rangle}$ and from $\pm 0.1\%$ to $\pm 1.4\%$ on $\frac{N_{ch}}{\meanmult}$. Furthermore, a 4Cx PYTHIA tune, which uses a different MPI model than the default configuration, was used as an input to observe the effect of different model assumptions used for the unfolding correction~\cite{bib:pythia:4Cx}. This affected the $\frac{N_{\varUpsilon}}{\langle N_{\varUpsilon} \rangle}$ by $\pm 0.2\%$ to $\pm 13.1\%$ and from $0\%$ to $\pm 1.9\%$ on $\frac{N_{ch}}{\meanmult}$.
The $\varUpsilon$ reconstruction efficiency vs. $N_{ch}$ did not show any change within uncertainties, so no dependence was assumed. The effect of this assumption was investigated by fitting the $\varUpsilon$ reconstruction efficiency vs multiplicity with a linear function and taking the $1\sigma$ upper limit as an input to the unfolding correction by correcting the $\varUpsilon$ yield. This changes the $\varUpsilon$ yield by $\pm 0.2\%$ to $\pm 3.8\%$.
The last contribution is due to the shape of the $N_{ch}$ distribution for minimum-bias events, which was checked by replacing the unfolded distribution with a fit using the negative binomial distribution and taking a $1\sigma$ upper and lower limit. It also enabled investigation of the possible influence of pile-up. This affects the $\varUpsilon$ yield by $\pm 0.3\%$ to $\pm 10.0\%$ and the $N_{ch}$ by $\pm 0.1\%$ to $\pm 2.7\%$.
The tracking efficiency mentioned previously also affects the $\varUpsilon$ yield vs. $N_{ch}$ and $N_{ch}$ itself by $\pm 0.4$ to $\pm 14\%$ and $\pm 3.5$ to $\pm 4.0\%$ respectively.
The total effect on $\frac{N_{\varUpsilon}}{\langle N_{\varUpsilon} \rangle}$ ranges from $\pm 6.1$ to $\pm 28.1\%$ and $\pm 4$ to $\pm 4.6\%$ on the $\frac{N_{ch}}{\meanmult}$.

\section{Results}
\label{sec:res}

We report a comprehensive measurement of the production of the $\varUpsilon$ states and show comparisons with corresponding production model calculations. This section is divided into subsections focusing on distinct measurements. First, the integrated production cross section and differential production cross sections vs. $p_{\mathrm{T}}$ and $y$ are shown and discussed. The next sub-section presents the cross section ratios. Finally, the charged particle multiplicity dependence is shown.

\subsection{$\varUpsilon$ production cross sections and spectra }

\begin{figure}[h!]
\begin{center}
	\includegraphics[width=0.6\textwidth]{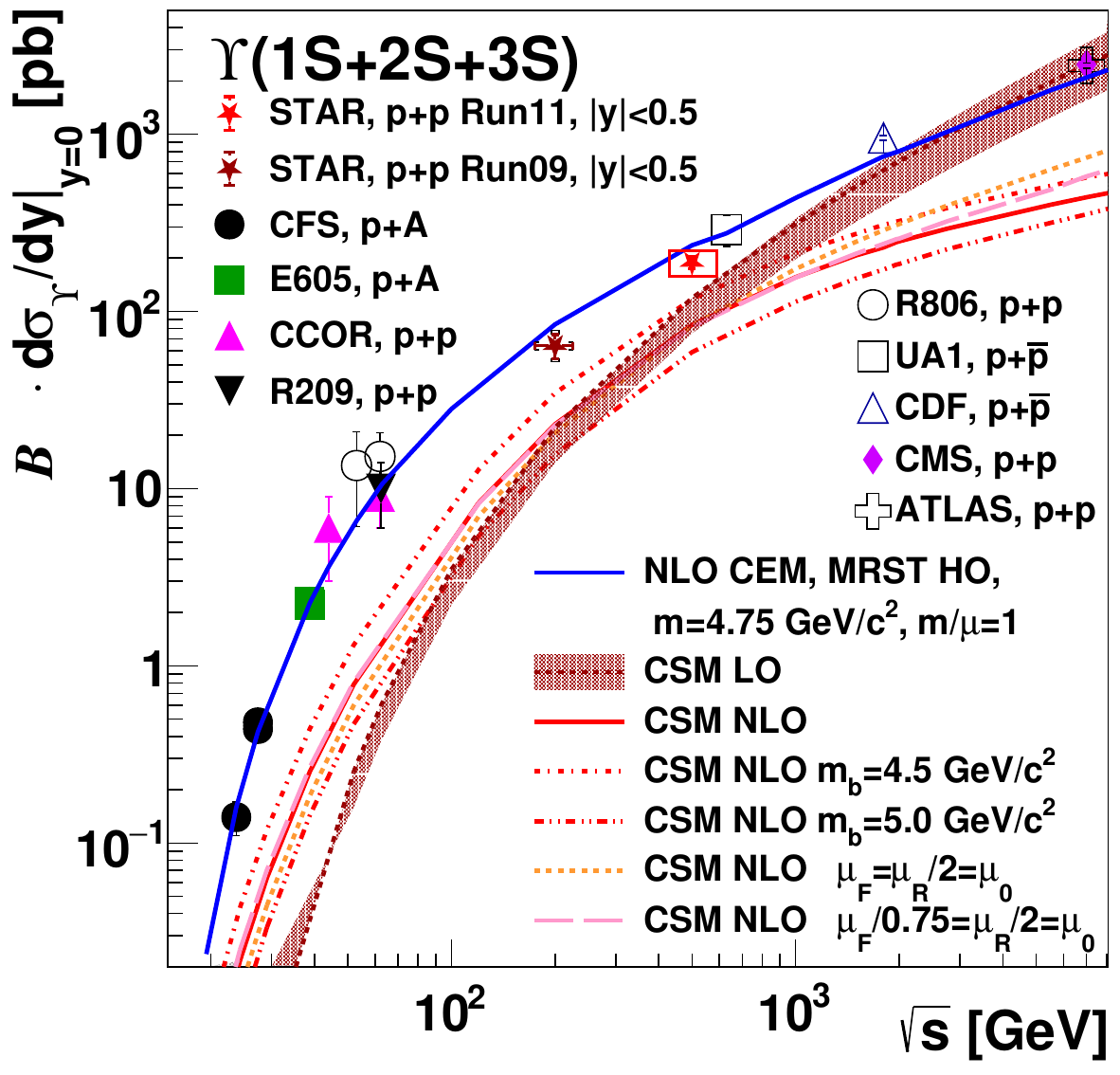}
\end{center}
\caption[$\varUpsilon$ integrated cross section] 
{ \label{Fig:Res:Int}  
	Integrated cross section of $\varUpsilon(1S+2S+3S)$ measured by STAR at $\sqrt{s}=200\:\mathrm{GeV}$~\cite{bib:Ups:STAR:dAu} and $\sqrt{s}=500\:\mathrm{GeV}$ compared to other experimental results~\cite{bib:Ups:AtlasRatio, bib:Ups:CMS:Xsec, bib:UpsCDF, bib:Ups:CFSpFe, bib:Ups:CFSppt, bib:Ups:CFSpp, bib:Ups:CFSpPtCu, bib:Ups:E605_pCu, bib:Ups:E605_pBe, bib:Ups:CCOR2, bib:Ups:CCOR2, bib:Ups:E866}, CEM calculation (blue line)~\cite{bib:Frawley2008} and CSM calculation at LO (red dotted line and band) and NLO (red lines: solid, dashed, dotted)~\cite{bib:lansberg:energy} plotted vs. center of mass energy.
}
\end{figure}

\begin{equation}
\label{Eq:Res:XsecPt}
\frac{B_{ee}}{2\pi p_{\mathrm{T}}} \frac{d\sigma^{2}_{\varUpsilon}}{dp_{\mathrm{T}}dy} = \frac{1}{2\pi p_{\mathrm{T}}}\frac{N_{\varUpsilon}(p_{\mathrm{T}})}{\Delta y \Delta p_{\mathrm{T}}} \frac{N_{ev}\mathcal{L}_{int}^{-1}}{\epsilon_{\varUpsilon}(p_{\mathrm{T}})}
\end{equation}

\begin{figure}[h!]
\begin{center}
	\subfloat[\label{Fig:Res:Spectra:Pt}]{%
	\includegraphics[width=0.49\textwidth]{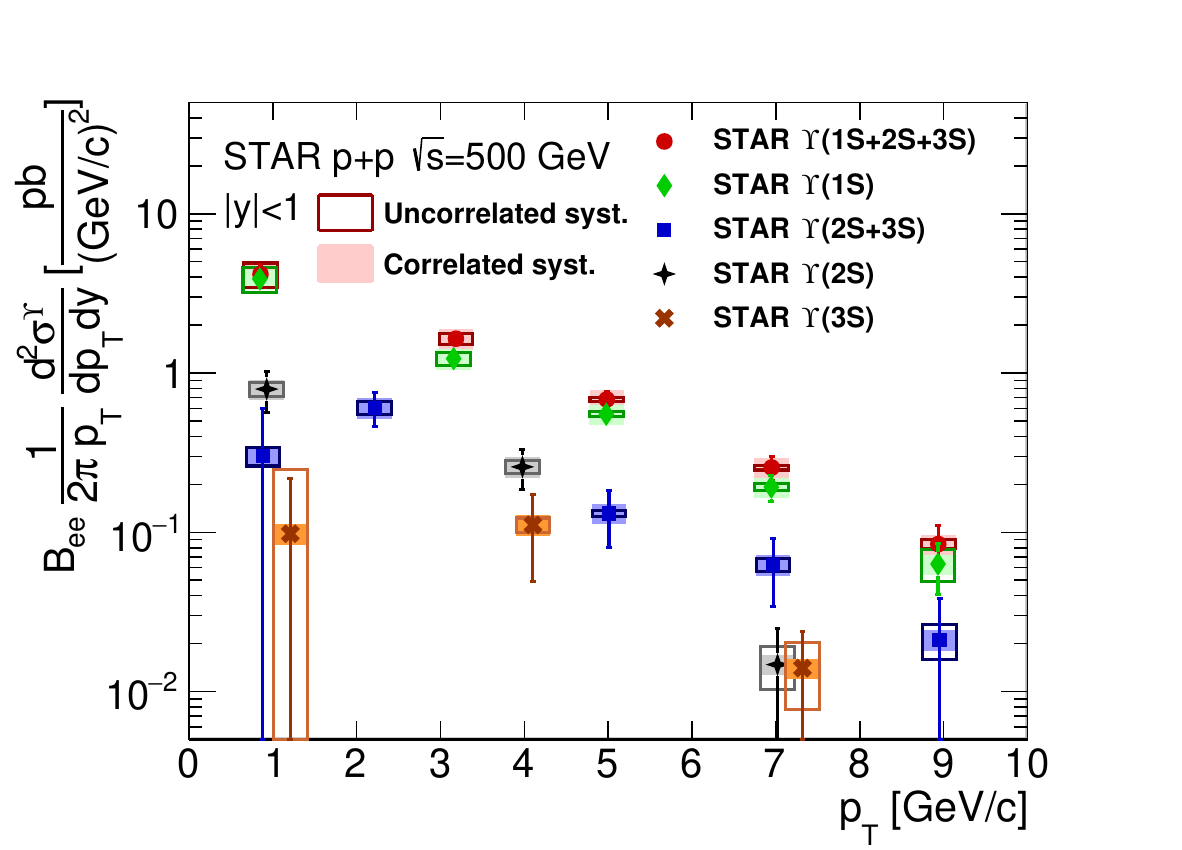}
		}
	\subfloat[\label{Fig:Res:Spectra:Rapidity}]{%
	\includegraphics[width=0.49\textwidth]{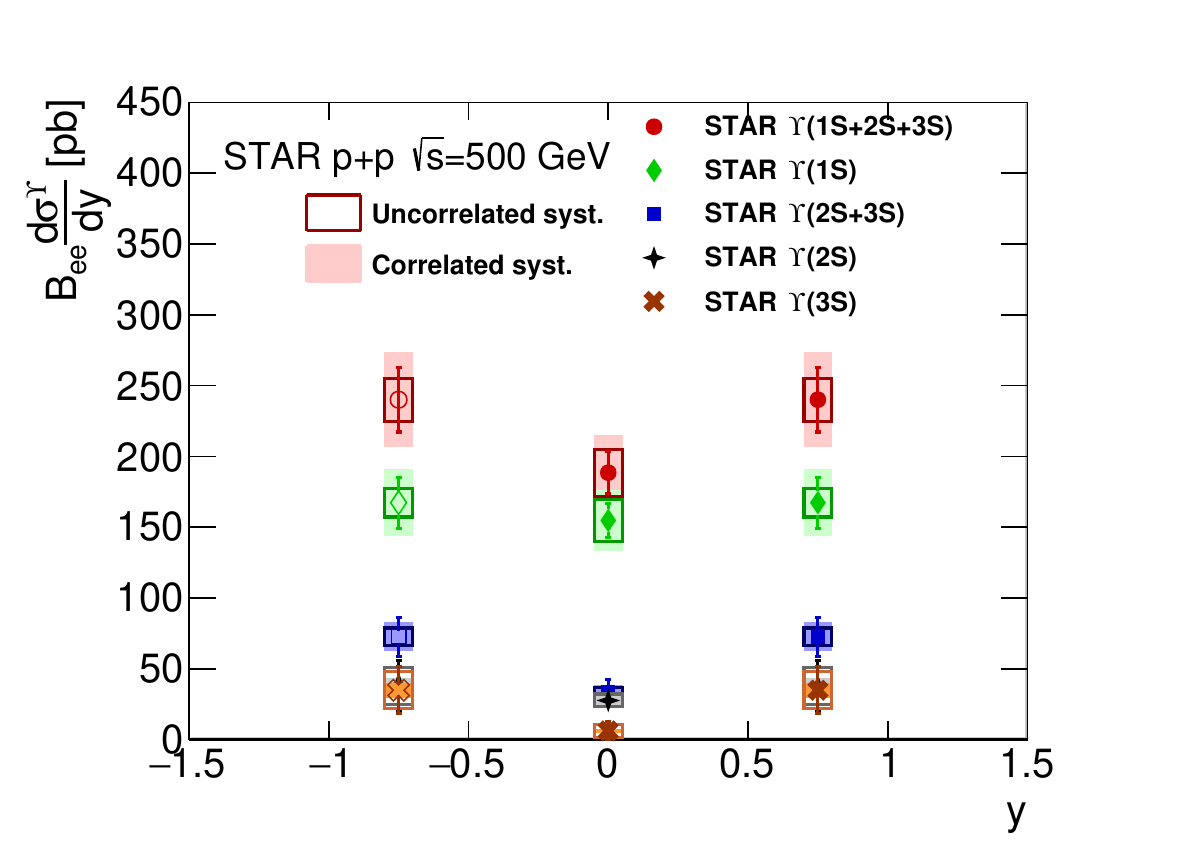}
		}
\end{center}
\caption[$\varUpsilon$ spectra] 
{ \label{Fig:Ups:Res:Spectra}  
(a) The $p_{\mathrm{T}}$ differential cross sections of $\varUpsilon(1S+2S+3S)$ (red circles), $\varUpsilon(2S+3S)$ (blue squares), $\varUpsilon(1S)$ (green diamonds), $\varUpsilon(2S)$ (black stars) and $\varUpsilon(3S)$ (brown crosses).
(b) Rapidity spectra for combined $\varUpsilon(1S+2S+3S)$ and each state separately (same as above). The hollow points at negative rapidity are mirror reflections of the forward rapidity data.
}
\end{figure} 

The differential cross section for $\varUpsilon$ production in \textit{$p+p$} collisions at $\sqrt{s}=500\:\mathrm{GeV}$ is calculated with Eq.~\ref{Eq:Res:XsecPt}. 
Here, $N_{\varUpsilon}$ is the measured $\varUpsilon$ yield, $\mathcal{L}_{int}$ is the integrated luminosity, $\epsilon_{\varUpsilon}$ is the $\varUpsilon$ reconstruction efficiency, $B_{ee}$ is the branching ratio to $e^{+}e^{-}$ and $N^{all}_{ev}$ is the total number of events, while $N_{ev}$ is the number of accepted events.
The $p_{T}$-integrated production cross section for combined $\varUpsilon(1S+2S+3S)$ states in $|y|<1$ is $199 \pm 13 (stat.) \pm 33 (syst.)\:\mathrm{pb}$. In the $|y|<0.5$ range, the cross section is $189 \pm 15 (stat.) \pm 31 (syst.)\:\mathrm{pb}$. The new STAR results are compared to the measurements at different energies~\cite{bib:Ups:AtlasRatio, bib:Ups:CMS:Xsec, bib:UpsCDF, bib:Ups:CFSpFe, bib:Ups:CFSppt, bib:Ups:CFSpp, bib:Ups:CFSpPtCu, bib:Ups:E605_pCu, bib:Ups:E605_pBe, bib:Ups:CCOR2, bib:Ups:CCOR2, bib:Ups:E866} in Fig.~\ref{Fig:Res:Int}. The data follow a similar trend as  a function of $\sqrt{s}$ and are consistent with the CEM calculation~\cite{bib:Frawley2008}, while the CSM calculation at leading-order (LO) (red dotted line and band) and next-to-leading-order (NLO) (red lines: solid, dashed, dotted)~\cite{bib:lansberg:energy} underestimate the data, especially below $1\:\mathrm{TeV}$.
It has to be noted that the CEM and CSM calculations include feed-down from $\varUpsilon(nS)$ states, while the contribution of $\chi_{bJ}$ states is excluded~\cite{bib:lansberg:private}. The NLO calculation is below the LO at high energy for scale choices $\mu_{R}>\mu_{F}$, as was identified in Ref.~\cite{bib:lansberg:curingNLO}. This can be resolved by fixing the factorisation scale or by re-summing specific corrections~\cite{bib:lansberg:matching}.

\begin{figure}[h!]
\begin{center}
	\begin{subfigure}{0.32\textwidth}
	\includegraphics[width=1.0\textwidth]{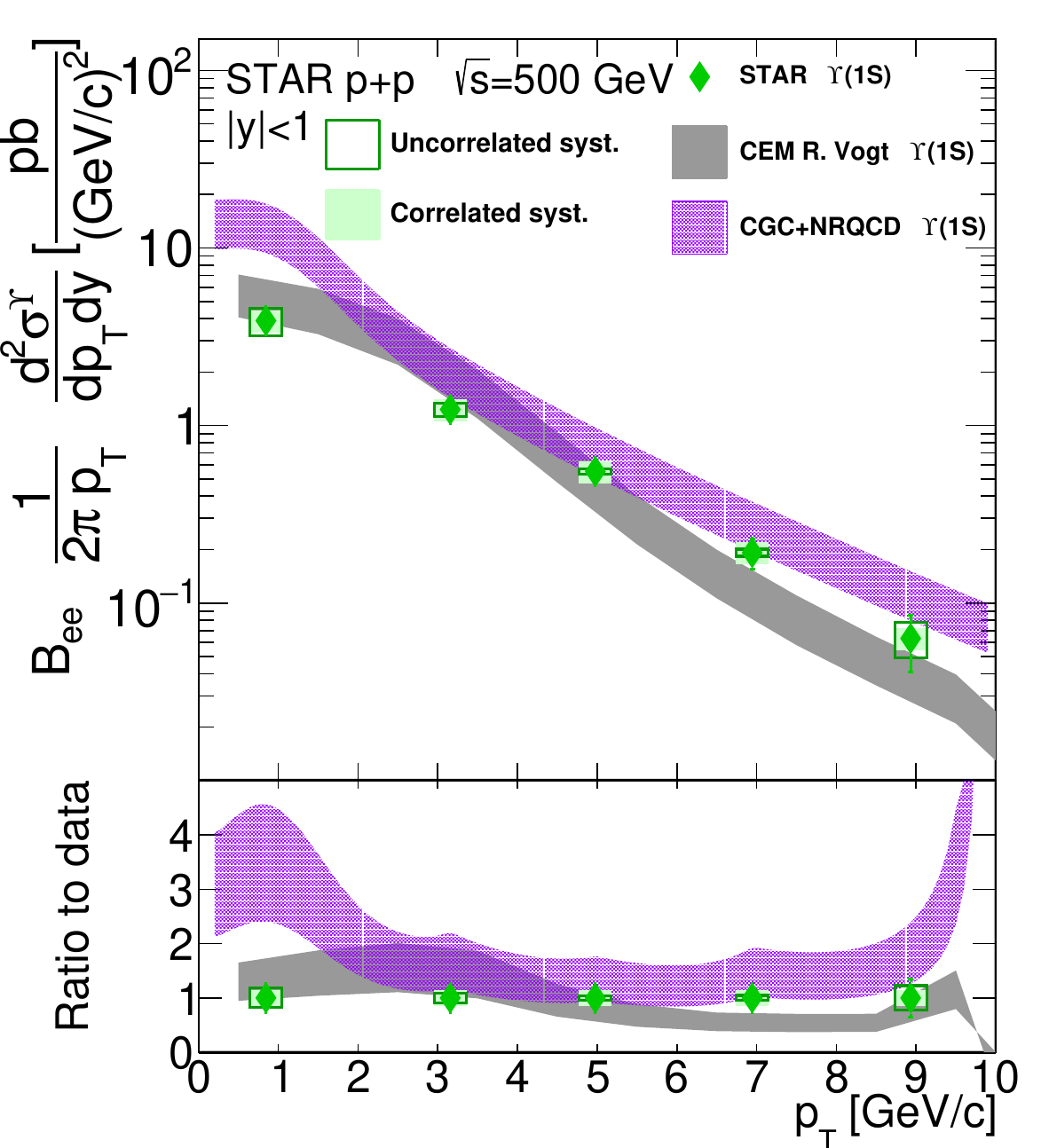}
		\caption[] { \label{Fig:Res:Spectra:Pt:1S}
		}
	\end{subfigure}
	\begin{subfigure}{0.32\textwidth}
	\includegraphics[width=1.0\textwidth]{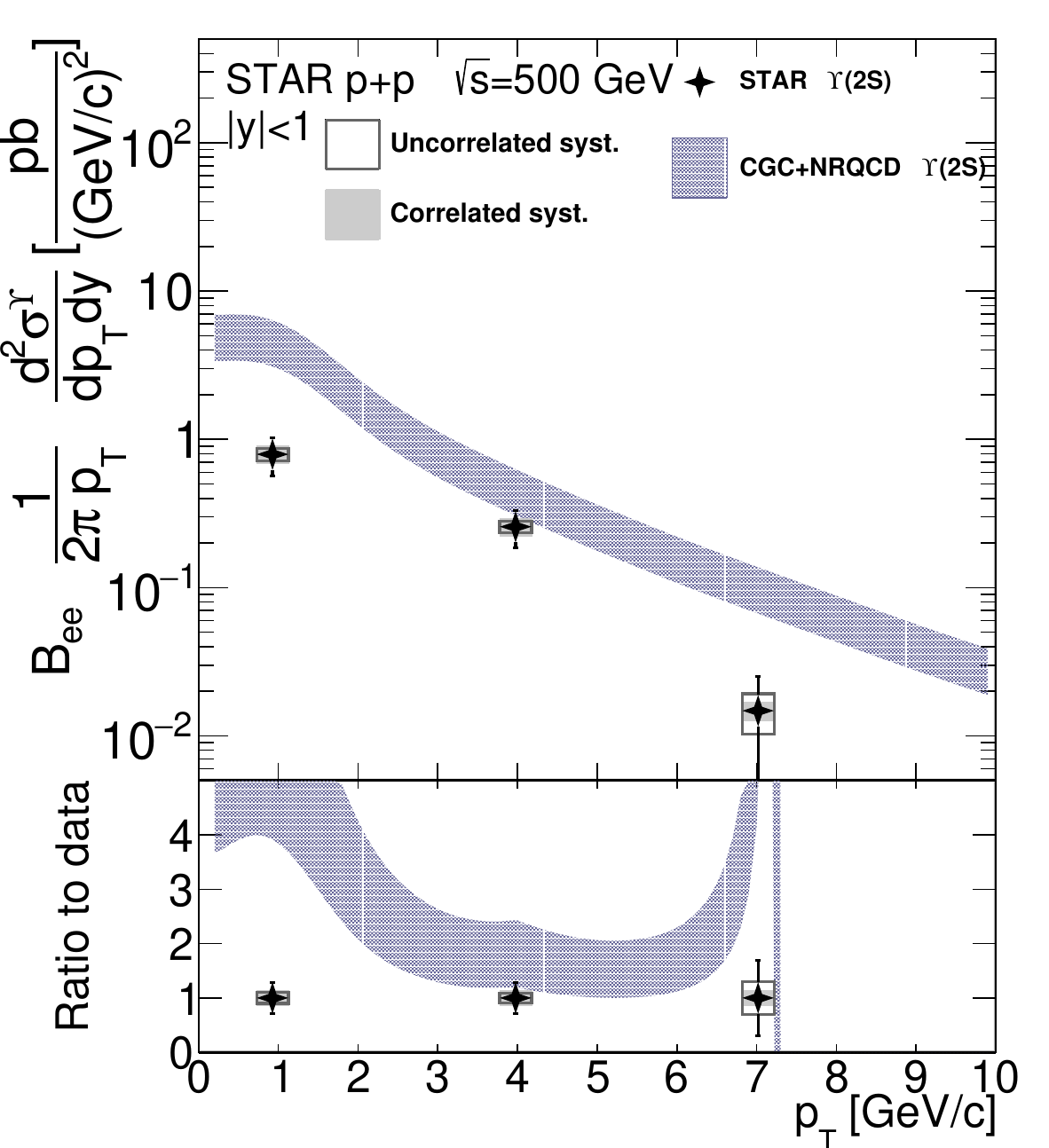}
		\caption[] { \label{Fig:Res:Spectra:Pt:2S}
		}
	\end{subfigure}
	\begin{subfigure}{0.32\textwidth}
	\includegraphics[width=1.0\textwidth]{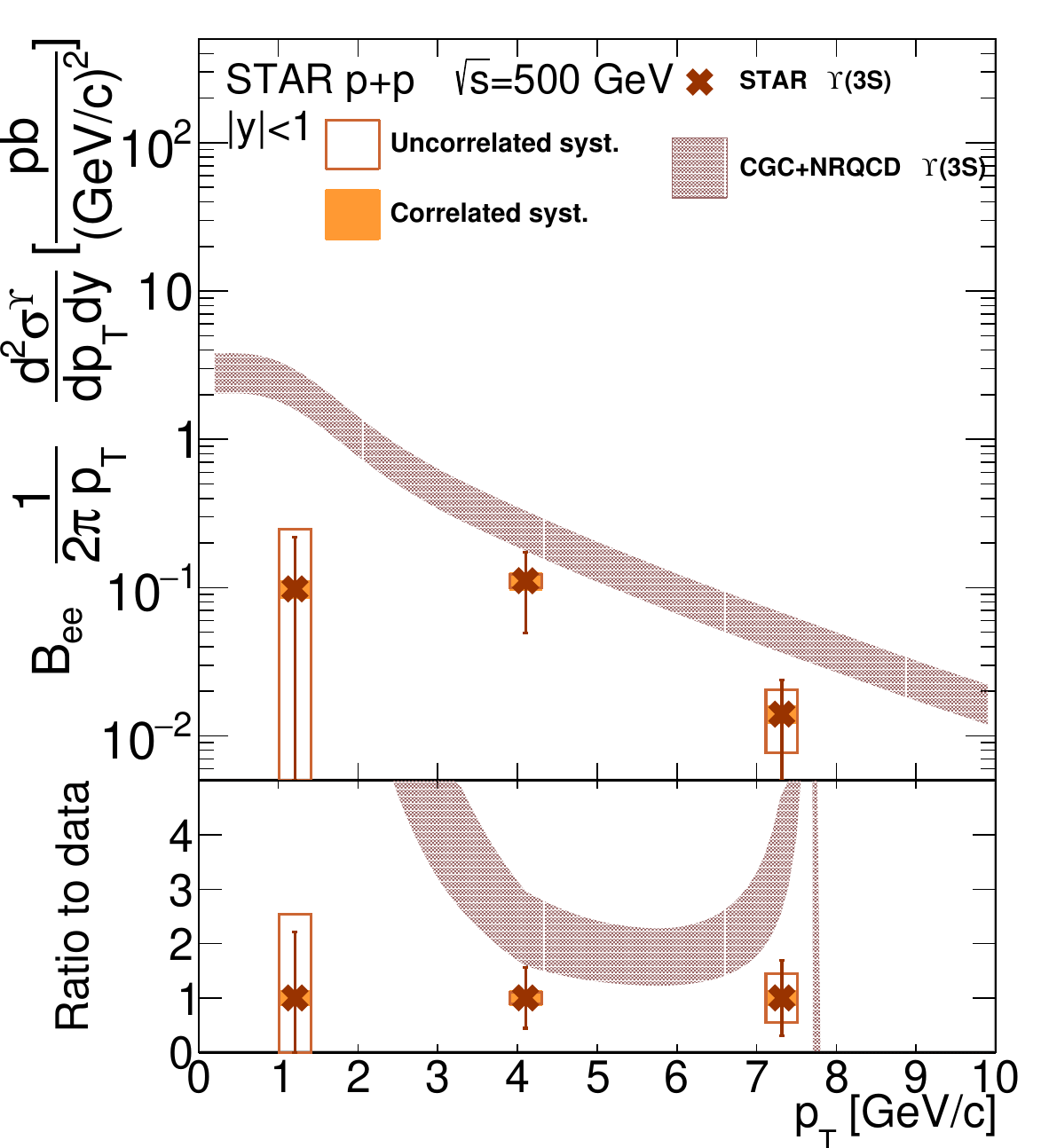}
		\caption[] { \label{Fig:Res:Spectra:Pt:3S}
		}
	\end{subfigure}
	
	%
\end{center}
\caption[$\varUpsilon$ $p_{\mathrm{T}}$ spectra] 
{ \label{Fig:Res:Spectra:Pt2}  
(a) The $\varUpsilon(1S)$ data are compared to the CEM calculation for inclusive $\varUpsilon(1S)$ (gray band)~\cite{bib:CEM_shadow,bib:vogt:private}. The results are also compared to a CGC+NRQCD calculation for direct $\varUpsilon(1S)$ (purple shaded band) ~\cite{bib:upsCGC,bib:jpsi_cgc,bib:YQMa}.
(b) Comparison of a CGC+NRQCD calculation for $\varUpsilon(2S)$ (light blue shaded band) to STAR data.
(c) STAR data comparison with a CGC+NRQCD calculation for $\varUpsilon(3S)$ (brown shaded band).
}
\end{figure} 

The differential cross sections vs. $p_{\mathrm{T}}$ and $y$ were also calculated and shown in Fig.~\ref{Fig:Res:Spectra:Pt} and Fig.~\ref{Fig:Res:Spectra:Rapidity}, respectively, for $\varUpsilon(1S+2S+3S)$ (red circles), $\varUpsilon(2S+3S)$ (blue squares), and for each $\varUpsilon$ states separately: $\varUpsilon(1S)$ (green diamonds), $\varUpsilon(2S)$ (black stars) and $\varUpsilon(3S)$ (brown crosses). 
The empty boxes indicate the uncorrelated systematic uncertainties, while the shaded areas correspond to the correlated (global) uncertainties.
The $\varUpsilon(2S)$ and $\varUpsilon(3S)$ yields are calculated in $3$ $p_{T}$ bins, while the rest are calculated in $5$ $p_{T}$, so are not directly comparable.
The $y$ dependence of $\varUpsilon(1S)$ and $\varUpsilon(2S)$ is flat in the measured range while, for $\varUpsilon(3S)$, a dip with $2\sigma$ significance is observed.
This is a downward fluctuation in the raw $\varUpsilon(3S)$ signal, as the reconstruction efficiency vs. $y$ for $\varUpsilon(3S)$ is similar to other states.
This fluctuation also contributes to the $\varUpsilon(1S+2S+3S)$ and $\varUpsilon(2S+3S)$ yields.

\begin{table}[h]
\centering
\begin{tabular}{|cccc|}
\hline
Model vs. $p_{T}$ & $\chi^{2}_{1S}/N_{DOF}$ & $\chi^{2}_{2S}/N_{DOF}$ & $\chi^{2}_{3S}/N_{DOF}$ \\
\hline
CGC+NRQCD & $10.7/5 = 2.15$ & $13.2/3 = 4.4$ & $14.9/3=4.98$ \\
\hline
CGC+NRQCD (1-st point removed) & $5.6/4 = 1.41$ & $6.1/2 = 3.07$ & $5.4/2 = 2.69$ \\
\hline
CEM & $4.8/5 = 0.95$ & - & - \\
\hline
	\end{tabular}
		\caption[Comparison of $\chi^{2}/N_{DOF}$ between measured $p_{\mathrm{T}}$ spectrum and model calculations.] 
		{ \label{Tab:Ana:Ups:pt:Chi2}
		Comparison of $\chi^{2}/N_{DOF}$ between measured $p_{\mathrm{T}}$ spectrum and model calculations.
		}
\end{table}

\begin{figure}[h!]
\begin{center}
	\begin{subfigure}{0.32\textwidth}
	\includegraphics[width=1.0\textwidth]{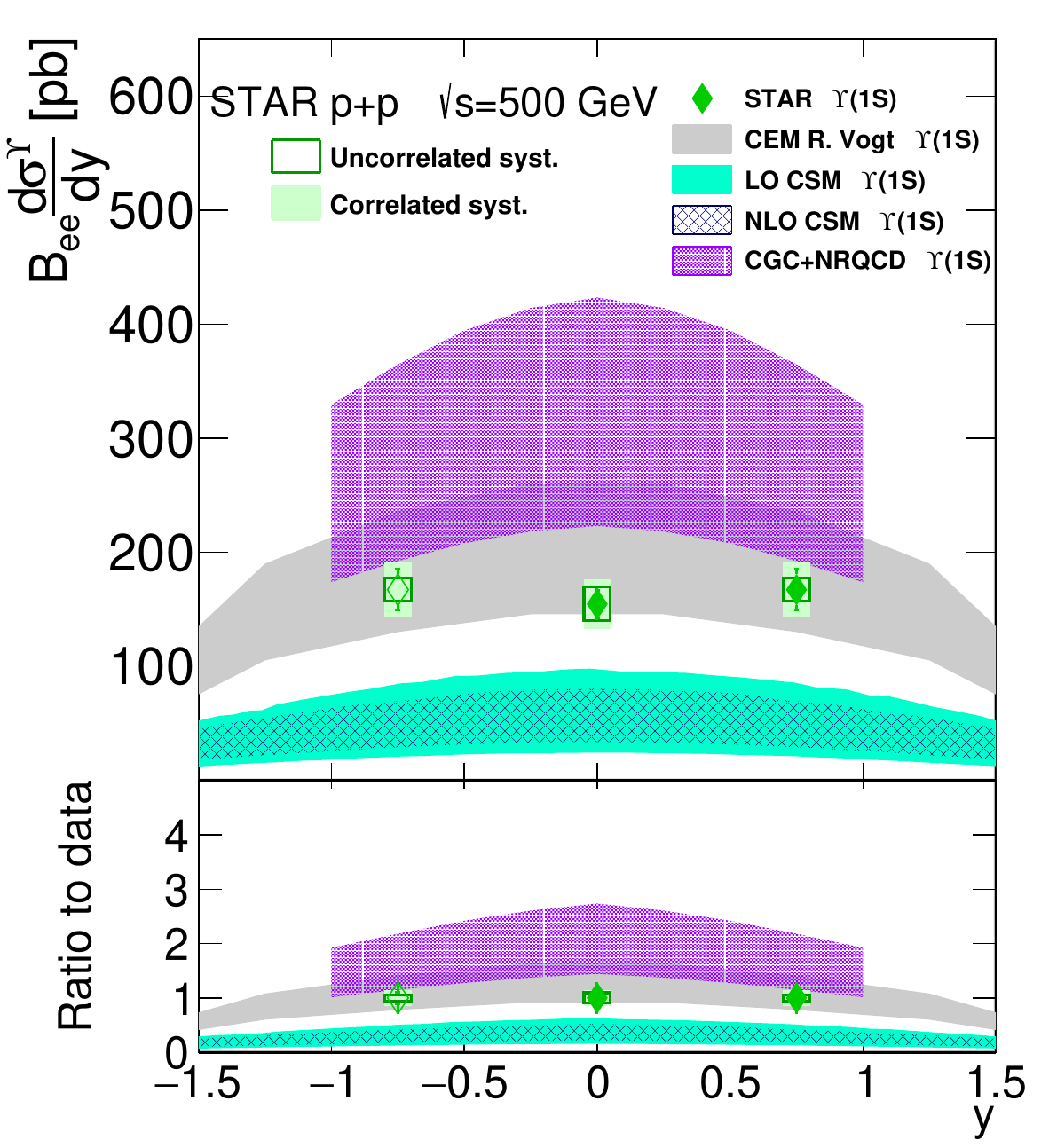}
		\caption[] { \label{Fig:Res:Spectra:Rap:1S}
		}
	\end{subfigure}
	\begin{subfigure}{0.32\textwidth}
	\includegraphics[width=1.0\textwidth]{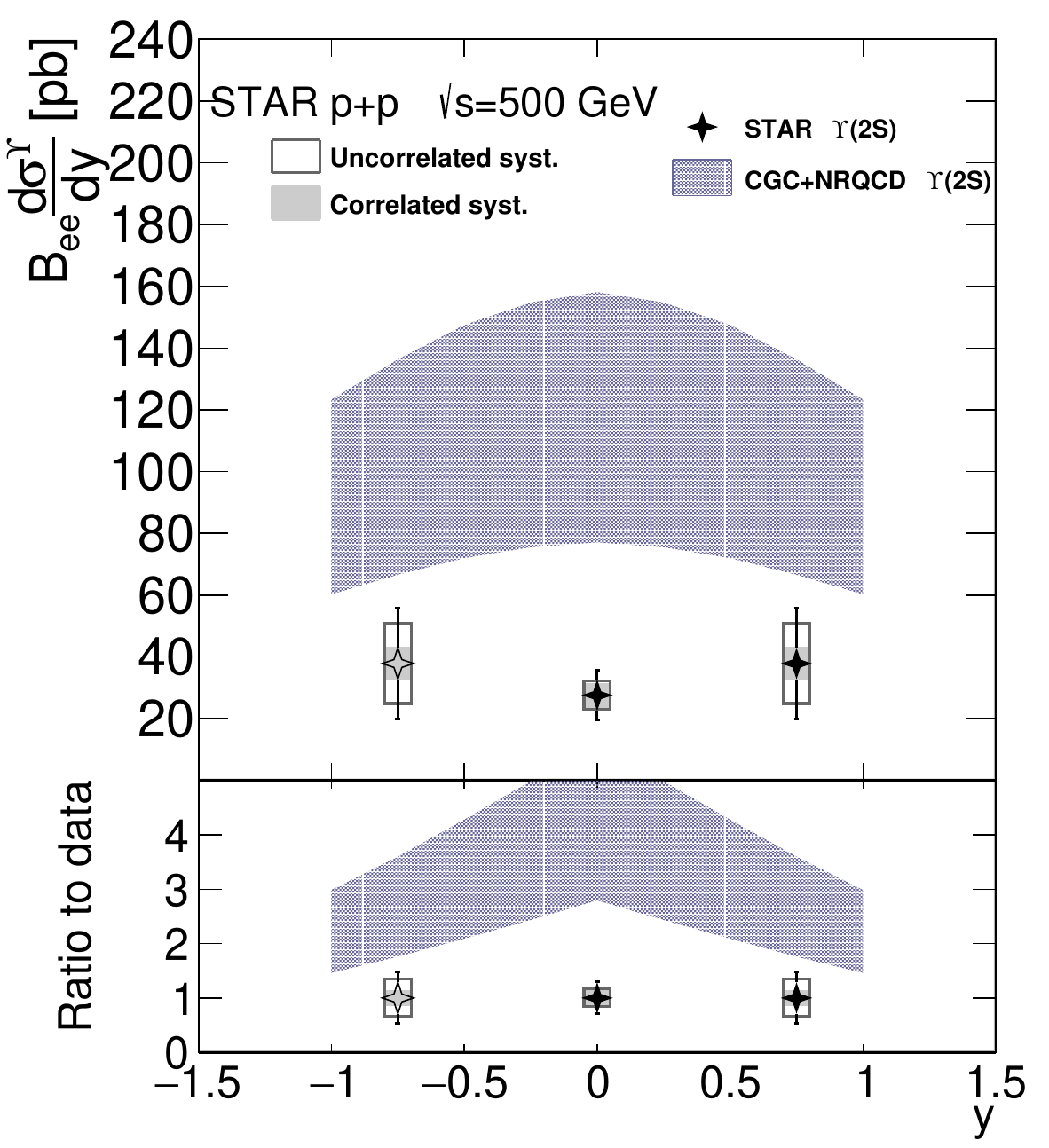}
		\caption[] { \label{Fig:Res:Spectra:Rap:2S}
		}
	\end{subfigure}
	\begin{subfigure}{0.32\textwidth}
	\includegraphics[width=1.0\textwidth]{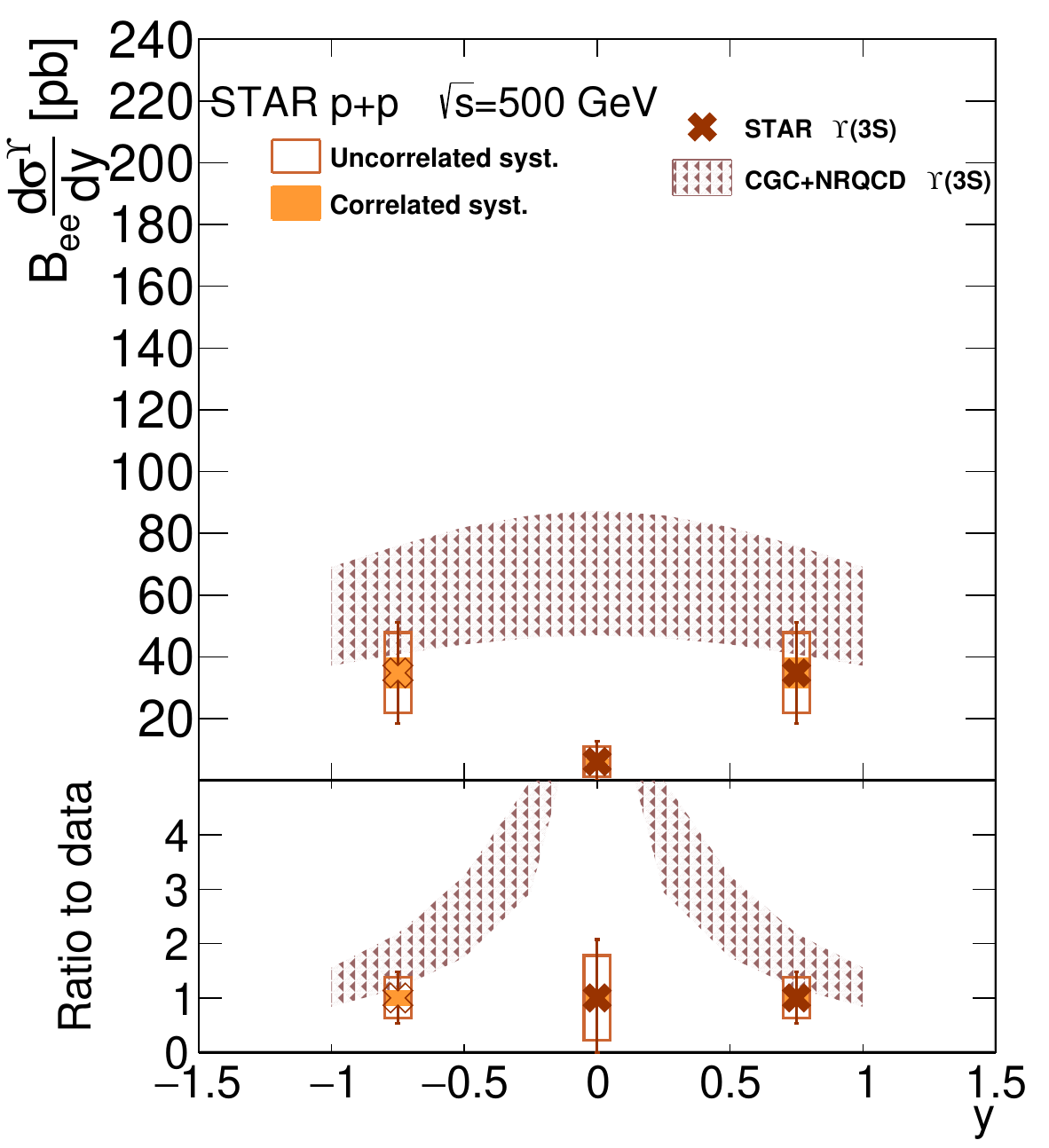}
		\caption[] { \label{Fig:Res:Spectra:Rap:3S}
		}
	\end{subfigure}
			
	\vspace{-0.4cm}

\end{center}
\caption[$\varUpsilon$ $y$ spectra] 
{ \label{Fig:Res:Spectra:Rap}  
(a) The $\varUpsilon(1S)$ data are compared to CEM calculation for inclusive $\varUpsilon(1S)$ (gray band)~\cite{bib:CEM_shadow,bib:vogt:private} and CGC+NRQCD predictions for direct $\varUpsilon(1S)$ (purple shaded band) ~\cite{bib:upsCGC,bib:jpsi_cgc,bib:YQMa} and Color Singlet model calculations at LO (teal band) and NLO (gray checked band)~\cite{bib:ups_csm}.
(b) Comparison of CGC+NRQCD calculation for $\varUpsilon(2S)$ (light blue shaded band) to STAR data.
(c) STAR data comparison to CGC+NRQCD calculation for $\varUpsilon(3S)$ (brown shaded band).
}
\end{figure}

The measured $p_{\mathrm{T}}$ spectra are compared to the CEM calculation for inclusive $\varUpsilon(1S)$ (gray band)~\cite{bib:CEM_shadow,bib:vogt:private} and CGC+NRQCD calculation for direct $\varUpsilon(1S)$ (purple shaded band) ~\cite{bib:upsCGC,bib:jpsi_cgc,bib:YQMa} in Fig.~\ref{Fig:Res:Spectra:Pt:1S}, while the STAR data are also compared to the CGC+NRQCD calculations for direct $\varUpsilon(2S)$ (brown shaded band) in Fig.~\ref{Fig:Res:Spectra:Pt:2S} and $\varUpsilon(3S)$ (light blue shaded band) in Fig.~\ref{Fig:Res:Spectra:Pt:3S}. As can be seen, the CEM calculation describes the $\varUpsilon(1S)$ cross section well, while the CGC+NRQCD calculation overestimates the data, especially for $p_{\mathrm{T}}<2\:\mathrm{GeV/c}$. The low-$p_{\mathrm{T}}$ part needs Sudakov resummation to improve the description, as indicated by the model authors~\cite{bib:venugopalan:private}. Feed-down effect is also not included for CGC+NRQCD, but would yield a $30$ to $55\%$ contribution to the total yield~\cite{bib:Upsilon:Overview:Andronic}. 
The model yields a good $\chi^{2}/N_{DOF}$ in the case of $\varUpsilon(1S)$ state and much larger values of $\chi^{2}/N_{DOF}$ for the excited states.
The fit, on the other hand, improves when the first point is excluded, as expected due to missing Sudakov resummation.
The $\chi^{2}_{1S}/N_{DOF}$ values are summarized in Tab.~\ref{Tab:Ana:Ups:pt:Chi2}.

\begin{table}[h]
\centering
\begin{tabular}{|cccc|}
\hline
Model vs. $y$ & $\chi^{2}_{1S}/N_{DOF}$ & $\chi^{2}_{2S}/N_{DOF}$ & $\chi^{2}_{3S}/N_{DOF}$ \\
\hline
CGC+NRQCD & $4.2/2 = 2.11$ & $7.2/2 = 3.6$ & $8.81/2=4.41$ \\
\hline
CEM & $0.62/2 = 0.31$ & - & - \\
\hline
CSM LO & $11.9/2 = 5.95$ & - & - \\
\hline
CSM NLO & $21.0/2 = 10.5$ & - & - \\
\hline
	\end{tabular}
		\caption[Comparison of $\chi^{2}/N_{DOF}$ between measured $y$ spectrum and model calculations.] 
		{ \label{Tab:Ana:Ups:rap:Chi2}
		Comparison of $\chi^{2}/N_{DOF}$ between measured $y$ spectrum and model calculations.
		}
\end{table}

\begin{figure}[h!]
\centering
		\includegraphics[width=0.75\textwidth]{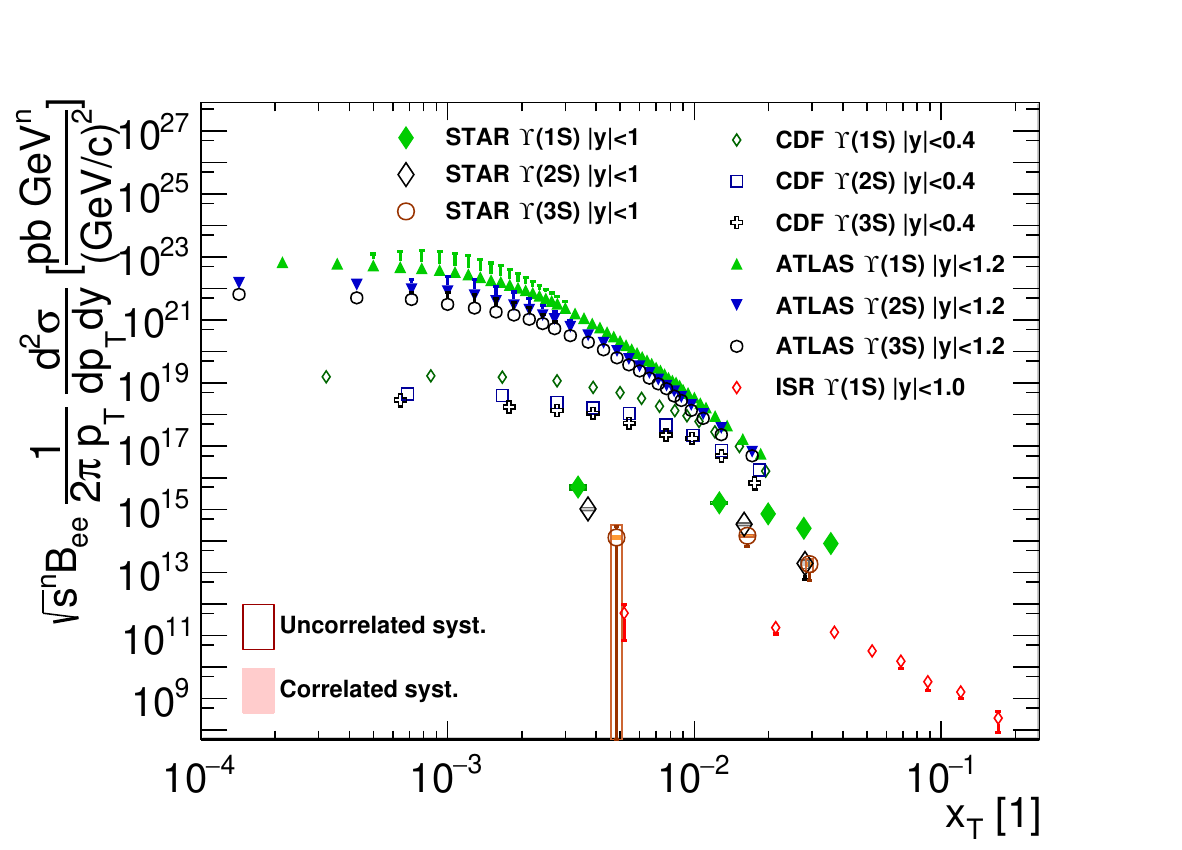}
	\caption[$\varUpsilon$ invariant cross section as a function of $x_{\mathrm{T}}$]
	{\label{Fig:Res:xt}
	$\varUpsilon$ invariant cross sections scaled with $\sqrt{s}^{n}$, where $n=5.6$ vs. $x_{\mathrm{T}}$ for $\varUpsilon(1S)$ (green closed diamonds), $\varUpsilon(2S)$ (black open diamonds) and $\varUpsilon(3S)$ (brown open circles) measured by STAR.
    The data are compared with $\varUpsilon(1S)$ results from Intersecting Storage Rings (ISR) (red open diamonds)~\cite{bib:UpsISR} and $\varUpsilon(1S)$, $\varUpsilon(2S)$, $\varUpsilon(3S)$ results measured by CDF (green open diamonds, blue open squares, black open crosses)~\cite{bib:UpsCDF} and ATLAS (green closed upward triangles, blue closed downward triangles, black open circles)~\cite{bib:Ups:AtlasRatio}. 
	}
\end{figure}

Rapidity spectra of $\varUpsilon(1S)$ are also compared in Fig.~\ref{Fig:Res:Spectra:Rap:1S} to the inclusive CEM (gray band)~\cite{bib:CEM_shadow,bib:vogt:private}, direct CSM at LO (teal band) and NLO (gray checked band)~\cite{bib:ups_csm} as well as direct CGC+NRQCD (purple shaded band) ~\cite{bib:upsCGC,bib:jpsi_cgc,bib:YQMa} calculations.
The Fig.~\ref{Fig:Res:Spectra:Rap:2S} compares the STAR data to direct CGC+NRQCD calculation for $\varUpsilon(2S)$ (light blue shaded band) while the comparison is extended to include  $\varUpsilon(3S)$ (brown shaded band) in Fig.~\ref{Fig:Res:Spectra:Rap:3S}. The rapidity spectra are well described by the CEM, but underestimated by both LO and NLO CSM calculations.
All $\varUpsilon(nS)$ cross sections at $y=0$ are overestimated by direct CGC+NRQCD calculation with large $\chi^{2}/N_{DOF}$. The directly produced $\varUpsilon$ constitute $45-70\%$~\cite{bib:Upsilon:Overview:Andronic} of the inclusive production, depending on $p_{\mathrm{T}}$, which would push the curves upwards.
The overestimation is due to the lack of aforementioned Sudakov resummation, which would improve the description at low $p_{\mathrm{T}}$~\cite{bib:venugopalan:private}.
In order to evaluate the quality of reproduction of the data by models, a comparison of $\chi^{2}/N_{DOF}$ values for each model and state is shown in Tab.~\ref{Tab:Ana:Ups:rap:Chi2}.

Finally, the $x_{\mathrm{T}}=\frac{2p_{\mathrm{T}}}{\sqrt{s}}$ distribution is calculated in order to investigate a possible $x_{\mathrm{T}}$ scaling of the $\varUpsilon$ production.
This is expected in pQCD~\cite{bib:pQCD:hipt}, where invariant cross sections ($\sigma^{inv}$) of hard processes at different collision energies should scale following the formula in Eq.~\ref{Eq:Intro:QCD:xt}, where the exponent $n$ is related to the number of partons participating in the hard process ~\cite{bib:pQCD:Kogut}, and $F(x_{\mathrm{T}})$ and $F'(x_{\mathrm{T}})$ are functions describing the $x_{\mathrm{T}}$ spectra.

\begin{equation}
\label{Eq:Intro:QCD:xt}
\sigma^{inv} \equiv E\frac{d^{3}\sigma}{d^{3}p} = \frac{F(x_{\mathrm{T}})}{p_{\mathrm{T}}^{n(x_{\mathrm{T}},\sqrt{s})}} = \frac{F'(x_{\mathrm{T}})}{\sqrt{s}^{n(x_{\mathrm{T}},\sqrt{s})}}
\end{equation}

For example, in a $2\rightarrow2$ process, $n=4$, and if there are more partons involved, then $n>4$. The $b\bar{b}$ pairs are produced in hard processes, but the bound state formation involves soft processes, which may break the scaling.
The scaled $x_{\mathrm{T}}$ distribution was calculated for each of the $\varUpsilon$ assuming $n=5.6$ as determined for $J/\psi$~\cite{bib:Jpsi:ZeboAuAu} and is shown in Fig.~\ref{Fig:Res:xt}. The STAR data are compared to results of other experiments~\cite{bib:UpsISR,bib:UpsCDF,bib:Ups:AtlasRatio,bib:Ups:CMS:diffXsec}. No scaling is observed within the currently measured $x_{\mathrm{T}}$ range, though more data are needed at high $x_{\mathrm{T}}$ values, where the distributions may potentially overlap. More data would show if the scaling takes place at high $x_{\mathrm{T}}$, so it motivates further future study.

\subsection{Cross section ratios}

\begin{figure}[h!]
\begin{center}
	\begin{subfigure}{0.49\textwidth}
	\includegraphics[width=1.0\textwidth]{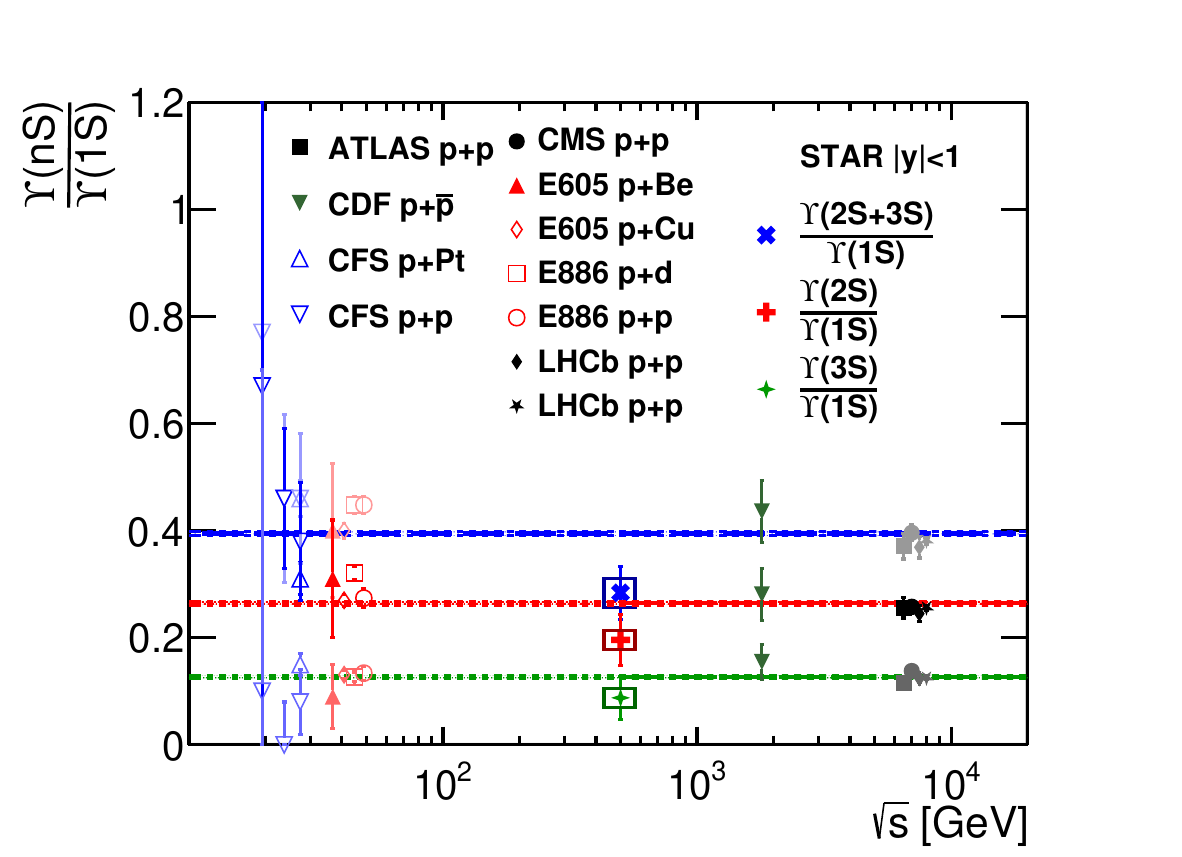}
		\caption[]{ \label{Fig:Res:Ratios:Sqrts}
		}
	\end{subfigure}	
	\begin{subfigure}{0.49\textwidth}
	\includegraphics[width=1.0\textwidth]{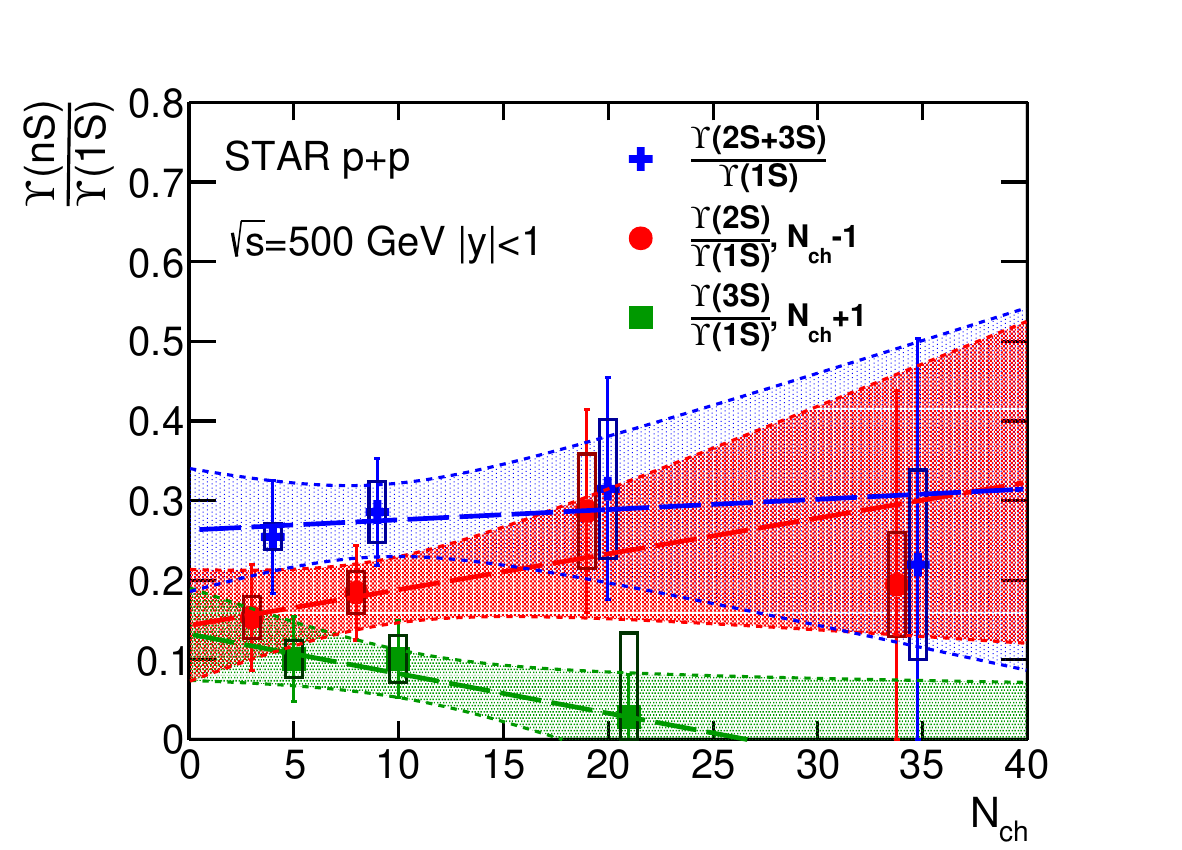}
		\caption[]{ \label{Fig:Res:Ratios:Mult}
		}
	\end{subfigure}
\end{center}
\caption[$\varUpsilon$ ratios] 
{ \label{Fig:Res:Ratios}  
(a) Cross section ratios of $\frac{\varUpsilon(nS)}{\varUpsilon(1S)}$ as a function of energy, where
the STAR measured ratios are $\frac{\varUpsilon(2S)}{\varUpsilon(1S)}$ (red cross), $\frac{\varUpsilon(3S)}{\varUpsilon(1S)}$ (green cross) and $\frac{\varUpsilon(2S+3S)}{\varUpsilon(1S)}$ (blue cross) compared to fits to the world data from~\cite{bib:Ups:Ratios} (red, green and blue lines respectively) with STAR data included.
The uncertainties on the fits are shown as bands around each line.
Measurements by other experiments in $p+\bar{p}$~\cite{bib:Ups:CDFratio}, \textit{$p+p$}~\cite{bib:Ups:CFSpp, bib:Ups:E866, bib:UpsCMS_2010, bib:Ups:AtlasRatio, bib:Ups:LHCbRatio, bib:Ups:LHCb8Tev} are also shown along with $p+A$~\cite{bib:Ups:E605_pBe, bib:Ups:E605_pCu, bib:Ups:E866, bib:Ups:CFSppt}.
(b) Dependence of $\frac{\varUpsilon(nS)}{\varUpsilon(1S)}$ cross section ratios on charged particle multiplicity. The STAR data for $\frac{\varUpsilon(2S+3S)}{\varUpsilon(1S)}$ (blue crosses), $\frac{\varUpsilon(2S)}{\varUpsilon(1S)}$ (red crosses) and $\frac{\varUpsilon(3S)}{\varUpsilon(1S)}$ (green crosses) are fitted with a linear function (blue, red and green lines).
The $\frac{\varUpsilon(2S)}{\varUpsilon(1S)}$ and $\frac{\varUpsilon(3S)}{\varUpsilon(1S)}$ data are shifted horizontally along $N_{ch}$ by -1 and +1 for clarity.
}
\end{figure} 

The cross section ratios are shown in Fig.~\ref{Fig:Res:Ratios:Sqrts} and are compared to the ratios measured by other experiments at different collision energies~\cite{bib:Ups:CFSpp, bib:Ups:CFSppt, bib:Ups:E605_pBe, bib:Ups:E605_pCu, bib:Ups:CDFratio, bib:UpsCMS_2010, bib:Ups:AtlasRatio, bib:Ups:LHCbRatio, bib:Ups:LHCb8Tev}.
The straight lines represent fits to $\varUpsilon$ states ratios measured by STAR as well as data taken from a systematic study of ratios~\cite{bib:Ups:Ratios}.
The ratios are not expected to depend much on collision energy, because the mass difference between the $\varUpsilon$ states is small.
The STAR data are below the fits with significance of $2.1\sigma$ for $\varUpsilon(2S+3S)/\varUpsilon(1S)$, $1.56\sigma$ for $\varUpsilon(2S)/\varUpsilon(1S)$ and $0.9\sigma$ for $\varUpsilon(3S)/\varUpsilon(1S)$.

\begin{figure}[h!]
\begin{center}
	\begin{subfigure}{0.32\textwidth}
	\includegraphics[width=1.0\textwidth]{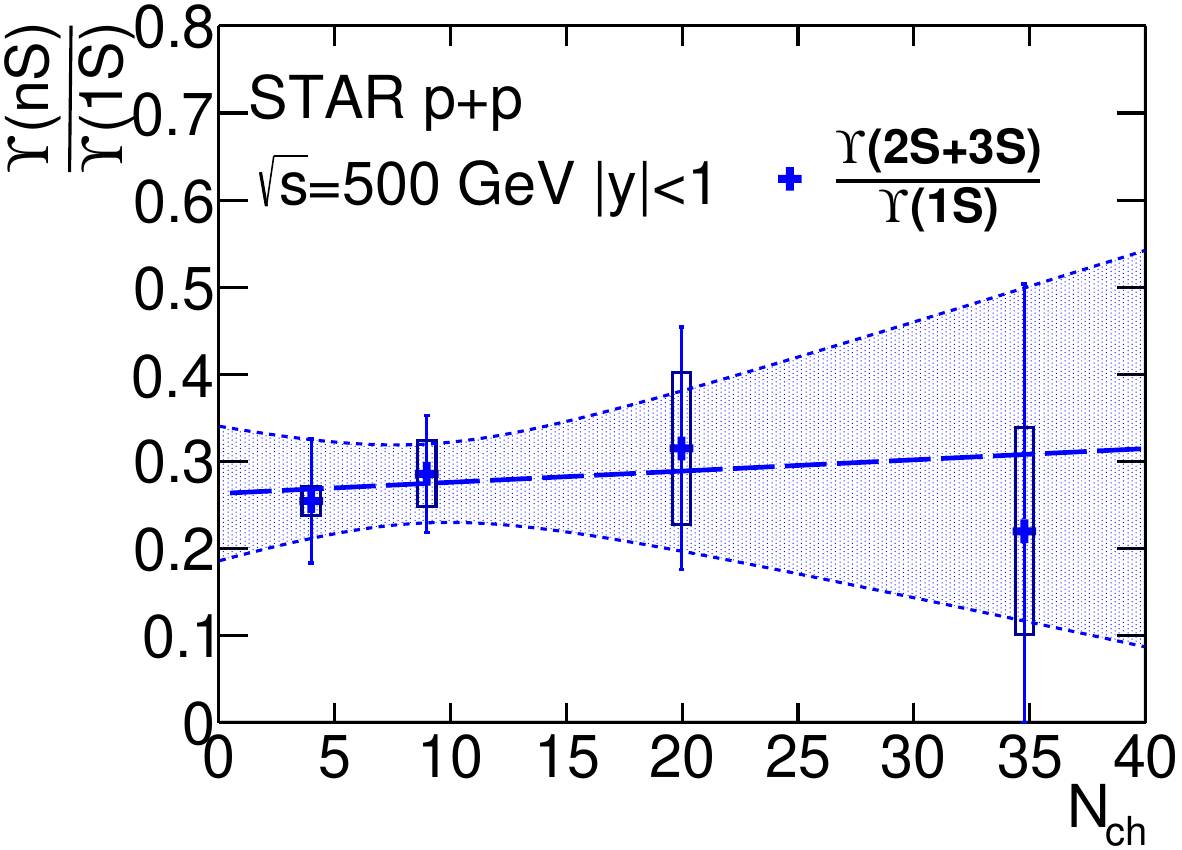}
		\caption[]{ \label{Fig:Res:Ratios:Mult:2S3Sto1S}
		}
	\end{subfigure}
	\begin{subfigure}{0.32\textwidth}
	\includegraphics[width=1.0\textwidth]{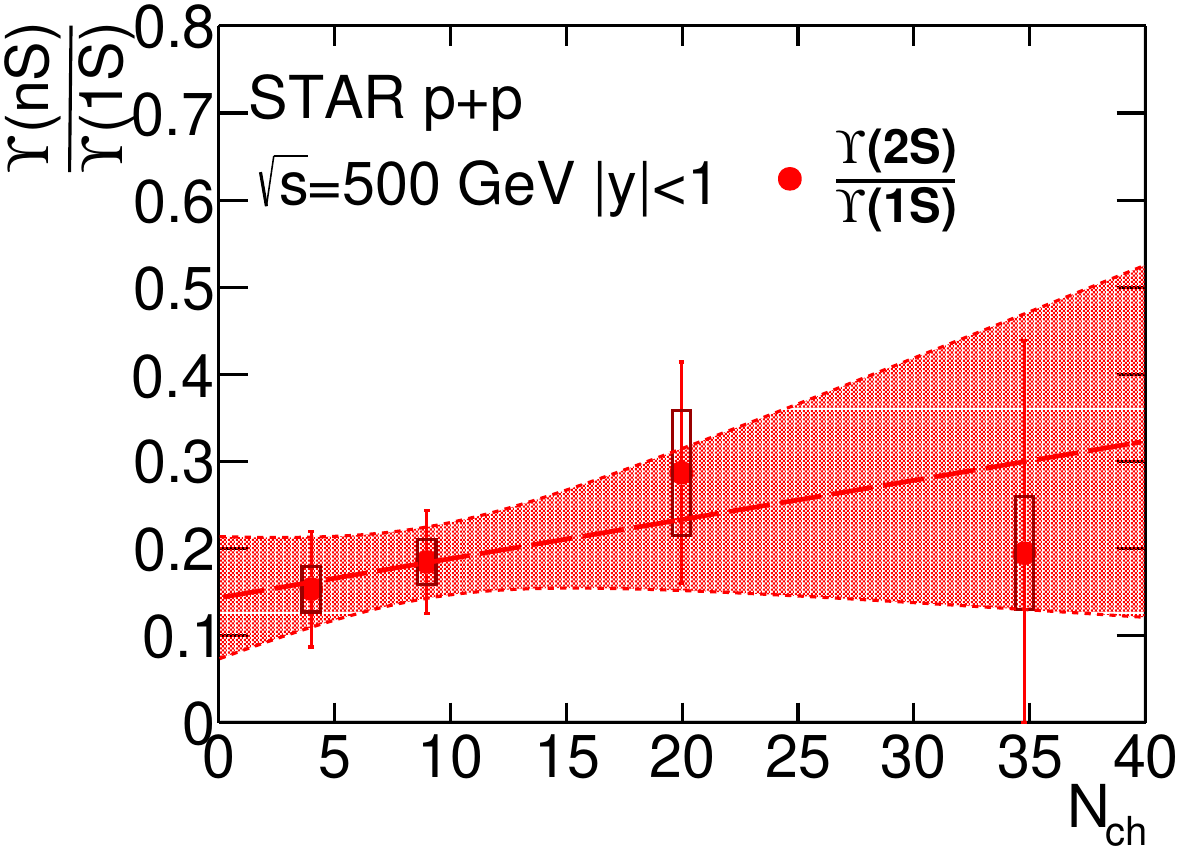}
		\caption[]{ \label{Fig:Res:Ratios:Mult:2Sto1S}
		}
	\end{subfigure}
	\begin{subfigure}{0.32\textwidth}
	\includegraphics[width=1.0\textwidth]{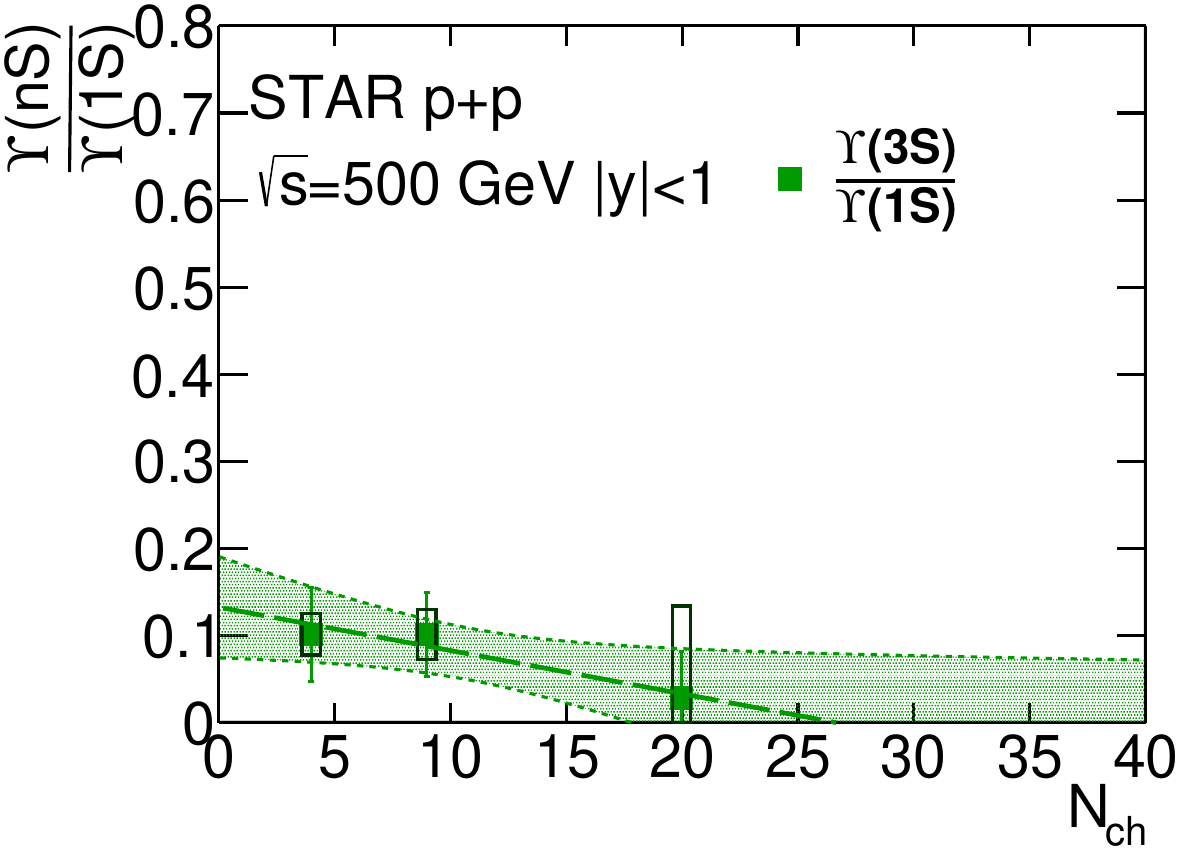}
		\caption[]{ \label{Fig:Res:Ratios:Mult:3Sto1S}
		}
	\end{subfigure}
\end{center}
\caption[$\varUpsilon$ ratios vs. $N_{ch}$] 
{ \label{Fig:Res:Ratios:1S2S3S}  
(a) Cross section ratio of $\varUpsilon(2S+3S)/\varUpsilon(1S)$ vs. $N_{ch}$.
(b) Cross section ratio of $\varUpsilon(2S)/\varUpsilon(1S)$ vs. $N_{ch}$.
(c) Cross section ratio of $\varUpsilon(3S)/\varUpsilon(1S)$ vs. $N_{ch}$.
}
\end{figure} 

The ratios are also studied vs. $N_{ch}$ and shown in Fig.~\ref{Fig:Res:Ratios:Mult}.
The points are placed in the corrected positions~\cite{bib:stat:dataPoints} based on the shape of $N_{ch}$ distribution of $\varUpsilon(1S)$, while the ratios $\varUpsilon(2S)/\varUpsilon(1S)$ and $\varUpsilon(3S)/\varUpsilon(1S)$ are shifted by $-1$ and $+1$, respectively for clarity.
In order to investigate the $N_{ch}$ dependence of the ratios, linear fits are performed (long dashed lines) and a $1\sigma$ uncertainty is also drawn (small dashed lines). The fits are shown separately for each ratio in Fig.~\ref{Fig:Res:Ratios:1S2S3S}.
Comover interactions are expected to cause a decrease of the ratios with $N_{ch}$~\cite{bib:Comover_FerreiroLansberg, bib:Comover_FerreiroLansberg_err}.
However, no significant dependence is observed within the uncertainties, which may indicate that the comover interaction plays only a small role for $\varUpsilon$.

\subsection{Charged particle multiplicity dependence}

\begin{figure}[h!]
\begin{center}
	\begin{subfigure}{0.49\textwidth}
	\includegraphics[width=1.0\textwidth]{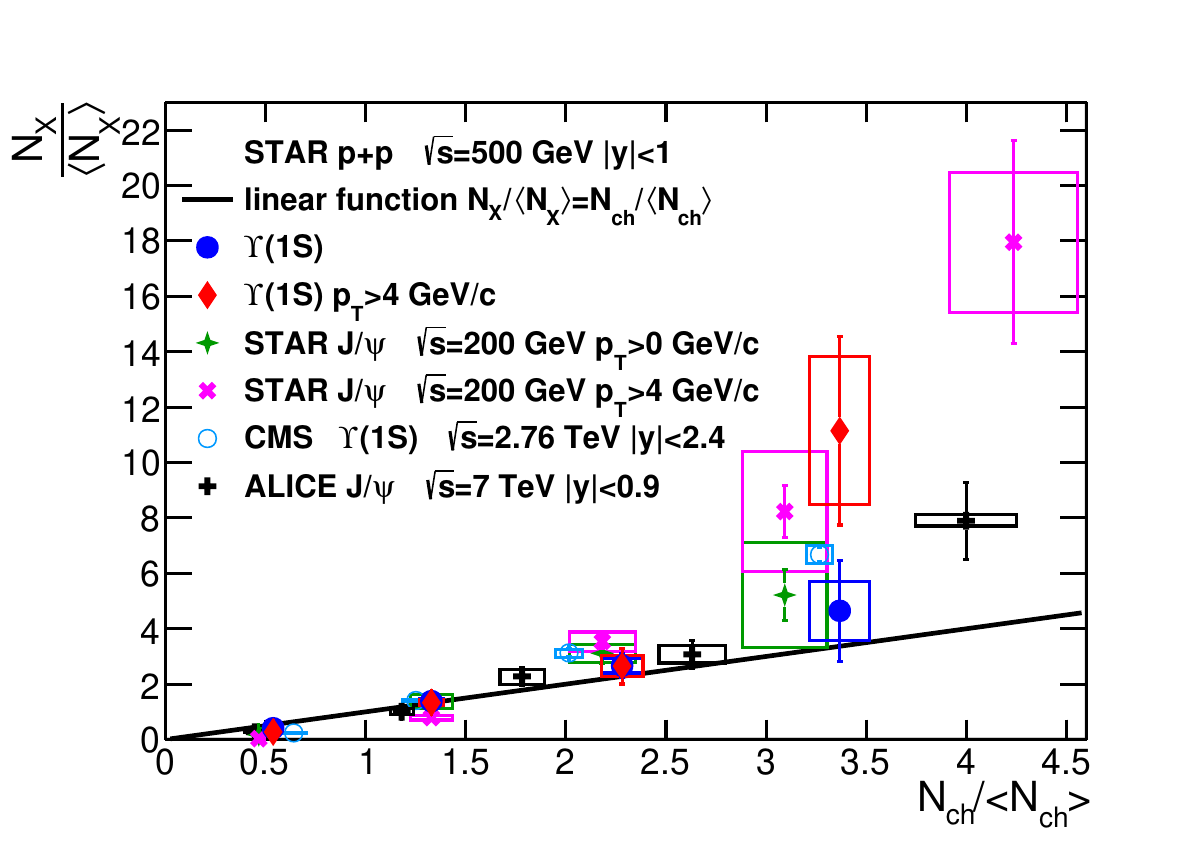}
		\caption[]{ \label{Fig:Res:EvAct:Data}
		}
	\end{subfigure}	
	\begin{subfigure}{0.49\textwidth}
	\includegraphics[width=1.0\textwidth]{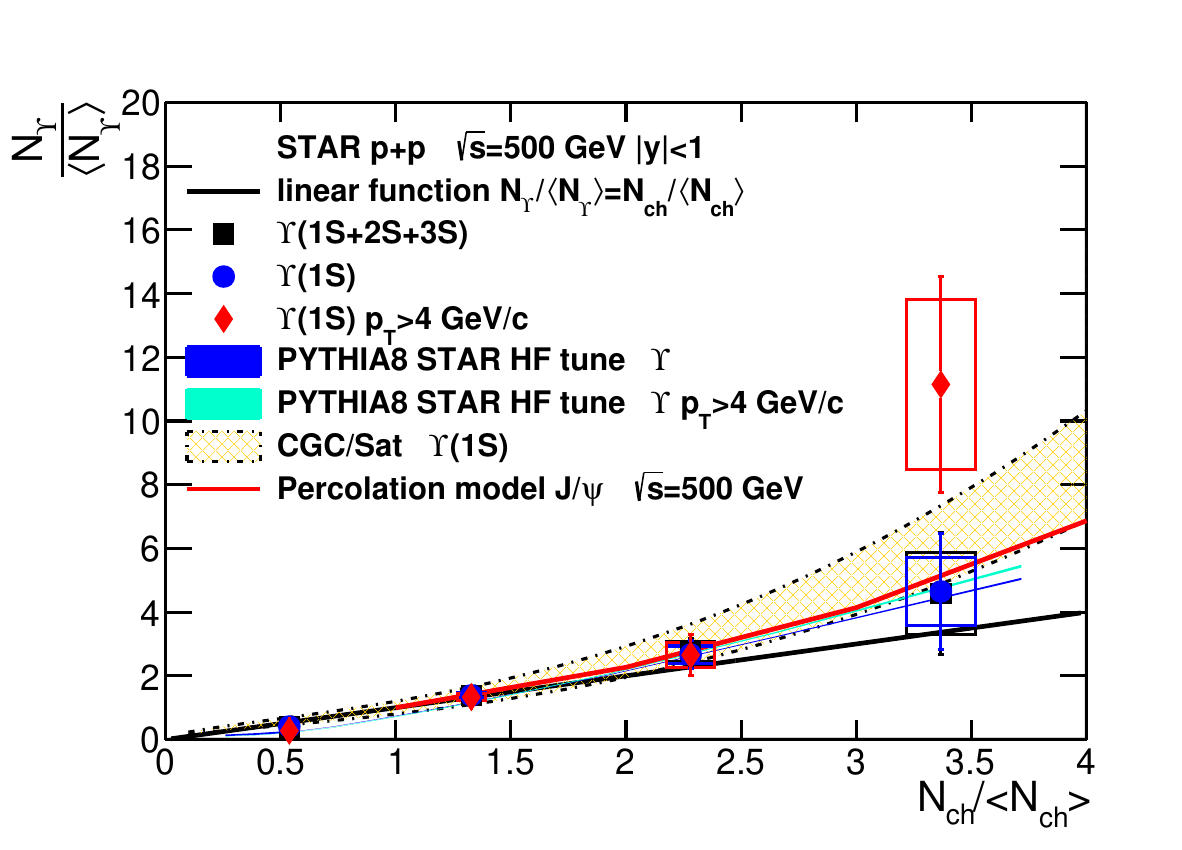}
		\caption[]{ \label{Fig:Res:EvAct:Models}
		}
	\end{subfigure}
\end{center}
\caption[$\varUpsilon$ multiplicity dependence] 
{ \label{Fig:Res:EvAct}  
(a) Yield as a function of $N_{ch}$ measured in \textit{$p+p$} collisions. STAR $\varUpsilon$ results at $\sqrt{s}=500\:\mathrm{GeV}$ are compared to $J/\psi$ at $\sqrt{s}=200\:\mathrm{GeV}$~\cite{bib:Jpsi:pp:STAR:mult} as well as $J/\psi$ ALICE~\cite{bib:ALICE:JpsiEventAct} and $\varUpsilon$ at CMS~\cite{bib:Ups:CMSactivity}.
(b) STAR $\varUpsilon$ results are compared to model calculations: PYTHIA8 with STAR Heavy Flavor Tune~\cite{bib:STAR_HFtune}, CGC-based Saturation model~\cite{bib:EventAct:Jpsi:3pom,bib:EventAct:QQ:CGC} and Percolation model for $J/\psi$~\cite{bib:PercolationJpsi}.
}
\end{figure}

The dependence of $\varUpsilon$ production on charged particle multiplicity is studied by calculating the yield $\frac{N_{\varUpsilon}}{\meanUps}$ vs. $\frac{N_{ch}}{\meanmult}$.
The signal yield is obtained from fits to the invariant mass spectra, the same method used earlier for the cross section calculations.
For the corrected $N_{ch}$ distribution, it is obtained from the low-luminosity minimum-bias data using unfolding and then corrected for the BBC trigger efficiency as explained in subsection~\ref{EffCorr}. The measured value of $\meanmult$ after the unfolding corrections is $\meanmult = 8.078 \pm 0.007$.
Bins are chosen as integer multiples of $\meanmult$ in a similar way to a previous STAR study~\cite{bib:Jpsi:pp:STAR:mult}, so that the bin limits are: $0-\meanmult$, $\meanmult-2\meanmult$, $2\meanmult-3\meanmult$ and $3\meanmult-8\meanmult$.
The STAR results of $\varUpsilon(1S+2S+3S)$ (black closed squares), $\varUpsilon(1S)$ (blue closed circles) and $\varUpsilon(1S)$ for $p_{\mathrm{T}}>4\:\mathrm{GeV/c}$ (red closed diamonds) are shown in Fig.~\ref{Fig:Res:EvAct:Data} and compared to $J/\psi$ measurements at STAR at $\sqrt{s}=200\:\mathrm{GeV}$~\cite{bib:Jpsi:pp:STAR:mult} and ALICE~\cite{bib:ALICE:JpsiEventAct} as well as $\varUpsilon$ measurements at CMS~\cite{bib:Ups:CMSactivity}. The STAR data for $p_{\mathrm{T}}>4\:\mathrm{GeV/c}$ follow the common trend of a strong increase with $N_{ch}$ observed at RHIC and the LHC both for $J/\psi$ and $\varUpsilon$. On the other hand, the $p_{\mathrm{T}}>0\:\mathrm{GeV/c}$ results are closer to a linear rise, with the notable exception of first point at low-$N_{ch}$.

The Fig.~\ref{Fig:Res:EvAct:Models} shows the $\frac{\varUpsilon}{\meanUps}$ measurements compared to the model calculations.
PYTHIA8 calculations using the STAR Heavy Flavor Tune~\cite{bib:STAR_HFtune} qualitatively reproduce the data trend. Similar behavior is exhibited by the CGC/Saturation model with a 3-pomeron contribution~\cite{bib:EventAct:Jpsi:3pom,bib:EventAct:QQ:CGC} to $\varUpsilon$ production. A Percolation Model calculation for $J/\psi$~\cite{bib:PercolationJpsi} is also shown and exhibits a strong increase as well, which is consistent with STAR $\varUpsilon$ results.

\section{Summary}
\label{sec:summ}

In this study we report the first measurement of the transverse momentum ($p_{\mathrm{T}}$) and rapidity ($y$) differential production cross sections for $\varUpsilon(1S)$, $\varUpsilon(2S)$, and $\varUpsilon(3S)$ states in \textit{$p+p$} collisions at $\sqrt{s}=500\:\mathrm{GeV}$.
The results are compared with various model predictions. While the inclusive $\varUpsilon(1S)$ spectra align well with the CEM~\cite{bib:CEM_shadow,bib:vogt:private}, the CGC+NRQCD~\cite{bib:upsCGC,bib:jpsi_cgc,bib:YQMa} calculations overestimate the spectra for all states, particularly at low $p_{\mathrm{T}}$.
The $y$ spectrum is also underestimated by the CSM~\cite{bib:ups_csm}, suggesting that these models may require updates to their long-distance matrix elements (LDMEs) based on the new STAR data.

We find no evidence of $x_{\mathrm{T}}$ scaling for $\varUpsilon$ states, and the cross section ratios are consistent with the world average within $2\sigma$. The charged particle multiplicity dependence of these ratios shows minimal influence on excited $\varUpsilon$ states, implying a minor effect from comover interactions.
The $\varUpsilon$ yield as a function of charged particle multiplicity follows the same trend observed for $J/\psi$ at RHIC and LHC energies, and is qualitatively described by models such as PYTHIA8~\cite{bib:STAR_HFtune}, CGC/Saturation~\cite{bib:EventAct:Jpsi:3pom,bib:EventAct:QQ:CGC}, and the String Percolation Model~\cite{bib:PercolationJpsi}.

These measurements impose new constraints on $\varUpsilon$ production models, particularly at low $p_{\mathrm{T}}$, and highlight the need for further studies to understand the interaction between $\varUpsilon$ production and soft QCD processes.
Future experiments at RHIC and the LHC will provide additional data to refine these models.

\section*{Acknowledgments}

We thank the RHIC Operations Group and RCF at BNL, the NERSC Center at LBNL, and the Open Science Grid consortium for providing resources and support.  This work was supported in part by the Office of Nuclear Physics within the U.S. DOE Office of Science, the U.S. National Science Foundation, National Natural Science Foundation of China, Chinese Academy of Science, the Ministry of Science and Technology of China and the Chinese Ministry of Education, the Higher Education Sprout Project by Ministry of Education at NCKU, the National Research Foundation of Korea, Czech Science Foundation and Ministry of Education, Youth and Sports of the Czech Republic, Hungarian National Research, Development and Innovation Office, New National Excellency Programme of the Hungarian Ministry of Human Capacities, Department of Atomic Energy and Department of Science and Technology of the Government of India, the National Science Centre and WUT ID-UB of Poland, the Ministry of Science, Education and Sports of the Republic of Croatia, German Bundesministerium f\"ur Bildung, Wissenschaft, Forschung and Technologie (BMBF), Helmholtz Association, Ministry of Education, Culture, Sports, Science, and Technology (MEXT), Japan Society for the Promotion of Science (JSPS) and Agencia Nacional de Investigaci\'on y Desarrollo (ANID) of Chile.


\bibliography{Bibliography}

\begin{thebibliography}{90}%
\makeatletter
\providecommand \@ifxundefined [1]{%
 \@ifx{#1\undefined}
}%
\providecommand \@ifnum [1]{%
 \ifnum #1\expandafter \@firstoftwo
 \else \expandafter \@secondoftwo
 \fi
}%
\providecommand \@ifx [1]{%
 \ifx #1\expandafter \@firstoftwo
 \else \expandafter \@secondoftwo
 \fi
}%
\providecommand \natexlab [1]{#1}%
\providecommand \enquote  [1]{``#1''}%
\providecommand \bibnamefont  [1]{#1}%
\providecommand \bibfnamefont [1]{#1}%
\providecommand \citenamefont [1]{#1}%
\providecommand \href@noop [0]{\@secondoftwo}%
\providecommand \href [0]{\begingroup \@sanitize@url \@href}%
\providecommand \@href[1]{\@@startlink{#1}\@@href}%
\providecommand \@@href[1]{\endgroup#1\@@endlink}%
\providecommand \@sanitize@url [0]{\catcode `\\12\catcode `\$12\catcode
  `\&12\catcode `\#12\catcode `\^12\catcode `\_12\catcode `\%12\relax}%
\providecommand \@@startlink[1]{}%
\providecommand \@@endlink[0]{}%
\providecommand \url  [0]{\begingroup\@sanitize@url \@url }%
\providecommand \@url [1]{\endgroup\@href {#1}{\urlprefix }}%
\providecommand \urlprefix  [0]{URL }%
\providecommand \Eprint [0]{\href }%
\providecommand \doibase [0]{https://doi.org/}%
\providecommand \selectlanguage [0]{\@gobble}%
\providecommand \bibinfo  [0]{\@secondoftwo}%
\providecommand \bibfield  [0]{\@secondoftwo}%
\providecommand \translation [1]{[#1]}%
\providecommand \BibitemOpen [0]{}%
\providecommand \bibitemStop [0]{}%
\providecommand \bibitemNoStop [0]{.\EOS\space}%
\providecommand \EOS [0]{\spacefactor3000\relax}%
\providecommand \BibitemShut  [1]{\csname bibitem#1\endcsname}%
\let\auto@bib@innerbib\@empty
\bibitem [{\citenamefont {Aidala}\ \emph {et~al.}(2018)\citenamefont {Aidala}
  \emph {et~al.}}]{bib:Jpsi:TSSA}%
  \BibitemOpen
  \bibfield  {author} {\bibinfo {author} {\bibfnamefont {C.}~\bibnamefont
  {Aidala}} \emph {et~al.} (\bibinfo {collaboration} {PHENIX Collaboration}),\
  }\href {https://doi.org/10.1103/physrevd.98.012006} {\bibfield  {journal}
  {\bibinfo  {journal} {Physical Review D}\ }\textbf {\bibinfo {volume} {98}},\
  \bibinfo {pages} {012006} (\bibinfo {year} {2018})}\BibitemShut {NoStop}%
\bibitem [{\citenamefont {Abdallah}\ \emph {et~al.}(2022)\citenamefont
  {Abdallah} \emph {et~al.}}]{bib:STAR:Jpsi:dAu:gluonic}%
  \BibitemOpen
  \bibfield  {author} {\bibinfo {author} {\bibfnamefont {M.~S.}\ \bibnamefont
  {Abdallah}} \emph {et~al.} (\bibinfo {collaboration} {STAR Collaboration}),\
  }\href {https://doi.org/10.1103/PhysRevLett.128.122303} {\bibfield  {journal}
  {\bibinfo  {journal} {Physical Review Letters}\ }\textbf {\bibinfo {volume}
  {128}},\ \bibinfo {pages} {122303} (\bibinfo {year} {2022})}\BibitemShut
  {NoStop}%
\bibitem [{\citenamefont {Chatrchyan}\ \emph {et~al.}(2012)\citenamefont
  {Chatrchyan} \emph {et~al.}}]{bib:Ups:CMS:Seq}%
  \BibitemOpen
  \bibfield  {author} {\bibinfo {author} {\bibfnamefont {S.}~\bibnamefont
  {Chatrchyan}} \emph {et~al.} (\bibinfo {collaboration} {CMS Collaboration}),\
  }\href {https://doi.org/10.1103/PhysRevLett.109.222301} {\bibfield  {journal}
  {\bibinfo  {journal} {Physical Review Letters}\ }\textbf {\bibinfo {volume}
  {109}},\ \bibinfo {pages} {222301} (\bibinfo {year} {2012})},\ \Eprint
  {https://arxiv.org/abs/1208.2826} {arXiv:1208.2826} \BibitemShut {NoStop}%
\bibitem [{\citenamefont {Matsui}\ and\ \citenamefont
  {Satz}(1986)}]{bib:Matsui1986}%
  \BibitemOpen
  \bibfield  {author} {\bibinfo {author} {\bibfnamefont {T.}~\bibnamefont
  {Matsui}}\ and\ \bibinfo {author} {\bibfnamefont {H.}~\bibnamefont {Satz}},\
  }\href {https://doi.org/10.1016/0370-2693(86)91404-8} {\bibfield  {journal}
  {\bibinfo  {journal} {Physics Letters B}\ }\textbf {\bibinfo {volume}
  {178}},\ \bibinfo {pages} {416} (\bibinfo {year} {1986})}\BibitemShut
  {NoStop}%
\bibitem [{\citenamefont {Politzer}(1973)}]{bib:pQCD:1}%
  \BibitemOpen
  \bibfield  {author} {\bibinfo {author} {\bibfnamefont {H.~D.}\ \bibnamefont
  {Politzer}},\ }\href {https://doi.org/10.1103/PhysRevLett.30.1346} {\bibfield
   {journal} {\bibinfo  {journal} {Physical Review Letters}\ }\textbf {\bibinfo
  {volume} {30}},\ \bibinfo {pages} {1346} (\bibinfo {year}
  {1973})}\BibitemShut {NoStop}%
\bibitem [{\citenamefont {Gross}\ and\ \citenamefont
  {Wilczek}(1973)}]{bib:pQCD:2}%
  \BibitemOpen
  \bibfield  {author} {\bibinfo {author} {\bibfnamefont {D.~J.}\ \bibnamefont
  {Gross}}\ and\ \bibinfo {author} {\bibfnamefont {F.}~\bibnamefont
  {Wilczek}},\ }\href {https://doi.org/10.1103/PhysRevLett.30.1343} {\bibfield
  {journal} {\bibinfo  {journal} {Physical Review Letters}\ }\textbf {\bibinfo
  {volume} {30}},\ \bibinfo {pages} {1343} (\bibinfo {year}
  {1973})}\BibitemShut {NoStop}%
\bibitem [{\citenamefont {Chao-Hsi}(1980)}]{bib:Onium:CS:1980}%
  \BibitemOpen
  \bibfield  {author} {\bibinfo {author} {\bibfnamefont {C.}~\bibnamefont
  {Chao-Hsi}},\ }\href {https://doi.org/10.1016/0550-3213(80)90175-3}
  {\bibfield  {journal} {\bibinfo  {journal} {Nuclear Physics B}\ }\textbf
  {\bibinfo {volume} {172}},\ \bibinfo {pages} {425} (\bibinfo {year}
  {1980})}\BibitemShut {NoStop}%
\bibitem [{\citenamefont {Berger}\ and\ \citenamefont
  {Jones}(1981)}]{bib:Onium:CS:JpsiUps}%
  \BibitemOpen
  \bibfield  {author} {\bibinfo {author} {\bibfnamefont {E.~L.}\ \bibnamefont
  {Berger}}\ and\ \bibinfo {author} {\bibfnamefont {D.}~\bibnamefont {Jones}},\
  }\href {https://doi.org/10.1103/PhysRevD.23.1521} {\bibfield  {journal}
  {\bibinfo  {journal} {Physical Review D}\ }\textbf {\bibinfo {volume} {23}},\
  \bibinfo {pages} {1521} (\bibinfo {year} {1981})}\BibitemShut {NoStop}%
\bibitem [{\citenamefont {Baier}\ and\ \citenamefont
  {R{\"{u}}ckl}(1981)}]{bib:Onium:CS:Baier}%
  \BibitemOpen
  \bibfield  {author} {\bibinfo {author} {\bibfnamefont {R.}~\bibnamefont
  {Baier}}\ and\ \bibinfo {author} {\bibfnamefont {R.}~\bibnamefont
  {R{\"{u}}ckl}},\ }\href {https://doi.org/10.1016/0370-2693(81)90636-5}
  {\bibfield  {journal} {\bibinfo  {journal} {Physics Letters B}\ }\textbf
  {\bibinfo {volume} {102}},\ \bibinfo {pages} {364} (\bibinfo {year}
  {1981})}\BibitemShut {NoStop}%
\bibitem [{\citenamefont {Bodwin}\ \emph {et~al.}(1995)\citenamefont {Bodwin},
  \citenamefont {Braaten},\ and\ \citenamefont {Lepage}}]{bib:Onium:CO:QCDana}%
  \BibitemOpen
  \bibfield  {author} {\bibinfo {author} {\bibfnamefont {G.~T.}\ \bibnamefont
  {Bodwin}}, \bibinfo {author} {\bibfnamefont {E.}~\bibnamefont {Braaten}},\
  and\ \bibinfo {author} {\bibfnamefont {G.~P.}\ \bibnamefont {Lepage}},\
  }\href {https://doi.org/10.1103/PhysRevD.51.1125} {\bibfield  {journal}
  {\bibinfo  {journal} {Physical Review D}\ }\textbf {\bibinfo {volume} {51}},\
  \bibinfo {pages} {1125} (\bibinfo {year} {1995})}\BibitemShut {NoStop}%
\bibitem [{\citenamefont {Bodwin}\ \emph {et~al.}(1997)\citenamefont {Bodwin},
  \citenamefont {Braaten},\ and\ \citenamefont
  {Lepage}}]{bib:Onium:CO:QCDanaErr}%
  \BibitemOpen
  \bibfield  {author} {\bibinfo {author} {\bibfnamefont {G.~T.}\ \bibnamefont
  {Bodwin}}, \bibinfo {author} {\bibfnamefont {E.}~\bibnamefont {Braaten}},\
  and\ \bibinfo {author} {\bibfnamefont {G.~P.}\ \bibnamefont {Lepage}},\
  }\href {https://doi.org/10.1103/PhysRevD.55.5853} {\bibfield  {journal}
  {\bibinfo  {journal} {Physical Review D}\ }\textbf {\bibinfo {volume} {55}},\
  \bibinfo {pages} {5853} (\bibinfo {year} {1997})}\BibitemShut {NoStop}%
\bibitem [{\citenamefont {Bodwin}\ \emph {et~al.}(1992)\citenamefont {Bodwin},
  \citenamefont {Braaten}, \citenamefont {Yuan},\ and\ \citenamefont
  {Lepage}}]{bib:Onium:CO:Bmeson}%
  \BibitemOpen
  \bibfield  {author} {\bibinfo {author} {\bibfnamefont {G.~T.}\ \bibnamefont
  {Bodwin}}, \bibinfo {author} {\bibfnamefont {E.}~\bibnamefont {Braaten}},
  \bibinfo {author} {\bibfnamefont {T.~C.}\ \bibnamefont {Yuan}},\ and\
  \bibinfo {author} {\bibfnamefont {G.~P.}\ \bibnamefont {Lepage}},\ }\href
  {https://doi.org/10.1103/PhysRevD.46.R3703} {\bibfield  {journal} {\bibinfo
  {journal} {Physical Review D}\ }\textbf {\bibinfo {volume} {46}},\ \bibinfo
  {pages} {R3703} (\bibinfo {year} {1992})}\BibitemShut {NoStop}%
\bibitem [{\citenamefont {Fritzsch}(1977)}]{bib:Jpsi:CEM:Fritzsch}%
  \BibitemOpen
  \bibfield  {author} {\bibinfo {author} {\bibfnamefont {H.}~\bibnamefont
  {Fritzsch}},\ }\href {https://doi.org/10.1016/0370-2693(77)90108-3}
  {\bibfield  {journal} {\bibinfo  {journal} {Physics Letters B}\ }\textbf
  {\bibinfo {volume} {67}},\ \bibinfo {pages} {217} (\bibinfo {year}
  {1977})}\BibitemShut {NoStop}%
\bibitem [{\citenamefont {Kosarzewski}(2022)}]{bib:Kosarzewski:2022enn}%
  \BibitemOpen
  \bibfield  {author} {\bibinfo {author} {\bibfnamefont {L.}~\bibnamefont
  {Kosarzewski}},\ }\href {https://doi.org/10.1088/1402-4896/ac6d1d} {\bibfield
   {journal} {\bibinfo  {journal} {Physica Scripta}\ }\textbf {\bibinfo
  {volume} {97}},\ \bibinfo {pages} {064004} (\bibinfo {year}
  {2022})}\BibitemShut {NoStop}%
\bibitem [{\citenamefont {Lansberg}(2020)}]{bib:lansberg_newObs}%
  \BibitemOpen
  \bibfield  {author} {\bibinfo {author} {\bibfnamefont {J.-P.}\ \bibnamefont
  {Lansberg}},\ }\href
  {https://doi.org/https://doi.org/10.1016/j.physrep.2020.08.007} {\bibfield
  {journal} {\bibinfo  {journal} {Physics Reports}\ }\textbf {\bibinfo {volume}
  {889}},\ \bibinfo {pages} {1} (\bibinfo {year} {2020})}\BibitemShut {NoStop}%
\bibitem [{\citenamefont {Acosta}\ \emph {et~al.}(2002)\citenamefont {Acosta}
  \emph {et~al.}}]{bib:UpsCDF}%
  \BibitemOpen
  \bibfield  {author} {\bibinfo {author} {\bibfnamefont {D.}~\bibnamefont
  {Acosta}} \emph {et~al.} (\bibinfo {collaboration} {CDF Collaboration}),\
  }\href {https://doi.org/10.1103/PhysRevLett.88.161802} {\bibfield  {journal}
  {\bibinfo  {journal} {Physical Review Letters}\ }\textbf {\bibinfo {volume}
  {88}},\ \bibinfo {pages} {161802} (\bibinfo {year} {2002})}\BibitemShut
  {NoStop}%
\bibitem [{\citenamefont {Acosta}\ \emph {et~al.}(2005)\citenamefont {Acosta}
  \emph {et~al.}}]{bib:UpsCDF_1960}%
  \BibitemOpen
  \bibfield  {author} {\bibinfo {author} {\bibfnamefont {D.}~\bibnamefont
  {Acosta}} \emph {et~al.} (\bibinfo {collaboration} {CDF Collaboration}),\
  }\href {https://doi.org/10.1103/PhysRevD.71.032001} {\bibfield  {journal}
  {\bibinfo  {journal} {Physical Review D}\ }\textbf {\bibinfo {volume} {71}},\
  \bibinfo {pages} {032001} (\bibinfo {year} {2005})}\BibitemShut {NoStop}%
\bibitem [{\citenamefont {Abe}\ \emph {et~al.}(1995)\citenamefont {Abe} \emph
  {et~al.}}]{bib:Ups:CDFratio}%
  \BibitemOpen
  \bibfield  {author} {\bibinfo {author} {\bibfnamefont {F.}~\bibnamefont
  {Abe}} \emph {et~al.} (\bibinfo {collaboration} {CDF Collaboration}),\ }\href
  {https://doi.org/10.1103/PhysRevLett.75.4358} {\bibfield  {journal} {\bibinfo
   {journal} {Physical Review Letters}\ }\textbf {\bibinfo {volume} {75}},\
  \bibinfo {pages} {4358} (\bibinfo {year} {1995})}\BibitemShut {NoStop}%
\bibitem [{\citenamefont {Acharya}\ \emph {et~al.}(2023)\citenamefont {Acharya}
  \emph {et~al.}}]{bib:ALICE_states}%
  \BibitemOpen
  \bibfield  {author} {\bibinfo {author} {\bibfnamefont {S.}~\bibnamefont
  {Acharya}} \emph {et~al.} (\bibinfo {collaboration} {ALICE Collaboration}),\
  }\href {https://doi.org/10.1140/epjc/s10052-022-10896-8} {\bibfield
  {journal} {\bibinfo  {journal} {The European Physical Journal C}\ }\textbf
  {\bibinfo {volume} {83}},\ \bibinfo {pages} {61} (\bibinfo {year}
  {2023})}\BibitemShut {NoStop}%
\bibitem [{\citenamefont {Sirunyan}\ \emph {et~al.}(2018)\citenamefont
  {Sirunyan} \emph {et~al.}}]{bib:Ups:CMSspectra}%
  \BibitemOpen
  \bibfield  {author} {\bibinfo {author} {\bibfnamefont {A.}~\bibnamefont
  {Sirunyan}} \emph {et~al.} (\bibinfo {collaboration} {CMS Collaboration}),\
  }\href {https://doi.org/https://doi.org/10.1016/j.physletb.2018.02.033}
  {\bibfield  {journal} {\bibinfo  {journal} {Physics Letters B}\ }\textbf
  {\bibinfo {volume} {780}},\ \bibinfo {pages} {251} (\bibinfo {year}
  {2018})}\BibitemShut {NoStop}%
\bibitem [{\citenamefont {Khachatryan}\ \emph {et~al.}(2011)\citenamefont
  {Khachatryan} \emph {et~al.}}]{bib:UpsCMS_2010}%
  \BibitemOpen
  \bibfield  {author} {\bibinfo {author} {\bibfnamefont {V.}~\bibnamefont
  {Khachatryan}} \emph {et~al.} (\bibinfo {collaboration} {CMS
  Collaboration}),\ }\href {https://doi.org/10.1103/PhysRevD.83.112004}
  {\bibfield  {journal} {\bibinfo  {journal} {Physical Review D}\ }\textbf
  {\bibinfo {volume} {83}},\ \bibinfo {pages} {112004} (\bibinfo {year}
  {2011})},\ \Eprint {https://arxiv.org/abs/1012.5545} {arXiv:1012.5545}
  \BibitemShut {NoStop}%
\bibitem [{\citenamefont {Khachatryan}\ \emph {et~al.}(2015)\citenamefont
  {Khachatryan} \emph {et~al.}}]{bib:Ups:CMS:diffXsec}%
  \BibitemOpen
  \bibfield  {author} {\bibinfo {author} {\bibfnamefont {V.}~\bibnamefont
  {Khachatryan}} \emph {et~al.} (\bibinfo {collaboration} {CMS
  Collaboration}),\ }\href {https://doi.org/10.1016/j.physletb.2015.07.037}
  {\bibfield  {journal} {\bibinfo  {journal} {Physics Letters B}\ }\textbf
  {\bibinfo {volume} {749}},\ \bibinfo {pages} {14} (\bibinfo {year} {2015})},\
  \Eprint {https://arxiv.org/abs/1501.07750} {arXiv:1501.07750} \BibitemShut
  {NoStop}%
\bibitem [{\citenamefont {Chatrchyan}\ \emph {et~al.}(2013)\citenamefont
  {Chatrchyan} \emph {et~al.}}]{bib:Ups:CMS:Xsec}%
  \BibitemOpen
  \bibfield  {author} {\bibinfo {author} {\bibfnamefont {S.}~\bibnamefont
  {Chatrchyan}} \emph {et~al.} (\bibinfo {collaboration} {CMS Collaboration}),\
  }\href {https://doi.org/10.1016/j.physletb.2013.10.033} {\bibfield  {journal}
  {\bibinfo  {journal} {Physics Letters B}\ }\textbf {\bibinfo {volume}
  {727}},\ \bibinfo {pages} {101} (\bibinfo {year} {2013})},\ \Eprint
  {https://arxiv.org/abs/1303.5900} {arXiv:1303.5900} \BibitemShut {NoStop}%
\bibitem [{\citenamefont {Aad}\ \emph {et~al.}(2013)\citenamefont {Aad} \emph
  {et~al.}}]{bib:Ups:AtlasRatio}%
  \BibitemOpen
  \bibfield  {author} {\bibinfo {author} {\bibfnamefont {G.}~\bibnamefont
  {Aad}} \emph {et~al.},\ }\href {https://doi.org/10.1103/PhysRevD.87.052004}
  {\bibfield  {journal} {\bibinfo  {journal} {Physical Review D}\ }\textbf
  {\bibinfo {volume} {87}},\ \bibinfo {pages} {052004} (\bibinfo {year}
  {2013})},\ \Eprint {https://arxiv.org/abs/1211.7255} {arXiv:1211.7255}
  \BibitemShut {NoStop}%
\bibitem [{\citenamefont {Aaij}\ \emph {et~al.}(2012)\citenamefont {Aaij} \emph
  {et~al.}}]{bib:Ups:LHCbRatio}%
  \BibitemOpen
  \bibfield  {author} {\bibinfo {author} {\bibfnamefont {R.}~\bibnamefont
  {Aaij}} \emph {et~al.} (\bibinfo {collaboration} {LHCb Collaboration}),\
  }\href {https://doi.org/10.1140/epjc/s10052-012-2025-y} {\bibfield  {journal}
  {\bibinfo  {journal} {The European Physical Journal C}\ }\textbf {\bibinfo
  {volume} {72}},\ \bibinfo {pages} {2025} (\bibinfo {year} {2012})},\ \Eprint
  {https://arxiv.org/abs/1202.6579} {arXiv:1202.6579} \BibitemShut {NoStop}%
\bibitem [{\citenamefont {Aaij}\ \emph {et~al.}(2013)\citenamefont {Aaij} \emph
  {et~al.}}]{bib:Ups:LHCb8Tev}%
  \BibitemOpen
  \bibfield  {author} {\bibinfo {author} {\bibfnamefont {R.}~\bibnamefont
  {Aaij}} \emph {et~al.} (\bibinfo {collaboration} {LHCb Collaboration}),\
  }\href {https://doi.org/10.1007/JHEP06(2013)064} {\bibfield  {journal}
  {\bibinfo  {journal} {Journal of High Energy Physics}\ }\textbf {\bibinfo
  {volume} {2013}},\ \bibinfo {pages} {64} (\bibinfo {year} {2013})},\ \Eprint
  {https://arxiv.org/abs/1304.6977} {arXiv:1304.6977} \BibitemShut {NoStop}%
\bibitem [{\citenamefont {Khachatryan}\ \emph {et~al.}(2017)\citenamefont
  {Khachatryan} \emph {et~al.}}]{bib:Ups:CMS:DPS}%
  \BibitemOpen
  \bibfield  {author} {\bibinfo {author} {\bibfnamefont {V.}~\bibnamefont
  {Khachatryan}} \emph {et~al.} (\bibinfo {collaboration} {CMS
  Collaboration}),\ }\href {https://doi.org/10.1007/jhep05(2017)013} {\bibfield
   {journal} {\bibinfo  {journal} {Journal of High Energy Physics}\ }\textbf
  {\bibinfo {volume} {2017}},\ \bibinfo {pages} {13} (\bibinfo {year}
  {2017})}\BibitemShut {NoStop}%
\bibitem [{\citenamefont {Sirunyan}\ \emph {et~al.}(2020)\citenamefont
  {Sirunyan} \emph {et~al.}}]{bib:Ups:CMS:DPS2}%
  \BibitemOpen
  \bibfield  {author} {\bibinfo {author} {\bibfnamefont {A.}~\bibnamefont
  {Sirunyan}} \emph {et~al.} (\bibinfo {collaboration} {CMS Collaboration}),\
  }\href {https://doi.org/https://doi.org/10.1016/j.physletb.2020.135578}
  {\bibfield  {journal} {\bibinfo  {journal} {Physics Letters B}\ }\textbf
  {\bibinfo {volume} {808}},\ \bibinfo {pages} {135578} (\bibinfo {year}
  {2020})}\BibitemShut {NoStop}%
\bibitem [{\citenamefont {Yoh}\ \emph {et~al.}(1978)\citenamefont {Yoh} \emph
  {et~al.}}]{bib:Ups:CFSpp}%
  \BibitemOpen
  \bibfield  {author} {\bibinfo {author} {\bibfnamefont {J.~K.}\ \bibnamefont
  {Yoh}} \emph {et~al.},\ }\href {https://doi.org/10.1103/PhysRevLett.41.684}
  {\bibfield  {journal} {\bibinfo  {journal} {Physical Review Letters}\
  }\textbf {\bibinfo {volume} {41}},\ \bibinfo {pages} {684} (\bibinfo {year}
  {1978})}\BibitemShut {NoStop}%
\bibitem [{\citenamefont {Kourkoumelis}\ \emph {et~al.}(1980)\citenamefont
  {Kourkoumelis} \emph {et~al.}}]{bib:UpsISR}%
  \BibitemOpen
  \bibfield  {author} {\bibinfo {author} {\bibfnamefont {C.}~\bibnamefont
  {Kourkoumelis}} \emph {et~al.},\ }\href
  {https://doi.org/10.1016/0370-2693(80)91024-2} {\bibfield  {journal}
  {\bibinfo  {journal} {Physics Letters B}\ }\textbf {\bibinfo {volume} {91}},\
  \bibinfo {pages} {481} (\bibinfo {year} {1980})}\BibitemShut {NoStop}%
\bibitem [{\citenamefont {Angelis}\ and\ \citenamefont
  {ohers}(1979)}]{bib:Ups:CCOR2}%
  \BibitemOpen
  \bibfield  {author} {\bibinfo {author} {\bibfnamefont {A.}~\bibnamefont
  {Angelis}}\ and\ \bibinfo {author} {\bibnamefont {ohers}},\ }\href
  {https://doi.org/10.1016/0370-2693(79)90563-X} {\bibfield  {journal}
  {\bibinfo  {journal} {Physics Letters B}\ }\textbf {\bibinfo {volume} {87}},\
  \bibinfo {pages} {398} (\bibinfo {year} {1979})}\BibitemShut {NoStop}%
\bibitem [{\citenamefont {Adare}\ \emph {et~al.}(2015)\citenamefont {Adare}
  \emph {et~al.}}]{bib:Ups:PHENIX:ppAuAu}%
  \BibitemOpen
  \bibfield  {author} {\bibinfo {author} {\bibfnamefont {A.}~\bibnamefont
  {Adare}} \emph {et~al.} (\bibinfo {collaboration} {PHENIX Collaboration}),\
  }\href {https://doi.org/10.1103/PhysRevC.91.024913} {\bibfield  {journal}
  {\bibinfo  {journal} {Physical Review C}\ }\textbf {\bibinfo {volume} {91}},\
  \bibinfo {pages} {024913} (\bibinfo {year} {2015})},\ \Eprint
  {https://arxiv.org/abs/1404.2246} {arXiv:1404.2246} \BibitemShut {NoStop}%
\bibitem [{\citenamefont {Abelev}\ \emph {et~al.}(2010)\citenamefont {Abelev}
  \emph {et~al.}}]{bib:Ups:STAR:pp}%
  \BibitemOpen
  \bibfield  {author} {\bibinfo {author} {\bibfnamefont {B.~I.}\ \bibnamefont
  {Abelev}} \emph {et~al.} (\bibinfo {collaboration} {STAR Collaboration}),\
  }\href {https://doi.org/10.1103/PhysRevD.82.012004} {\bibfield  {journal}
  {\bibinfo  {journal} {Physical Review D}\ }\textbf {\bibinfo {volume} {82}},\
  \bibinfo {pages} {012004} (\bibinfo {year} {2010})},\ \Eprint
  {https://arxiv.org/abs/1001.2745} {arXiv:1001.2745} \BibitemShut {NoStop}%
\bibitem [{\citenamefont {Adamczyk}\ \emph {et~al.}(2014)\citenamefont
  {Adamczyk} \emph {et~al.}}]{bib:Ups:STAR:dAu}%
  \BibitemOpen
  \bibfield  {author} {\bibinfo {author} {\bibfnamefont {L.}~\bibnamefont
  {Adamczyk}} \emph {et~al.} (\bibinfo {collaboration} {STAR Collaboration}),\
  }\href {https://doi.org/10.1016/j.physletb.2014.06.028} {\bibfield  {journal}
  {\bibinfo  {journal} {Physics Letters B}\ }\textbf {\bibinfo {volume}
  {735}},\ \bibinfo {pages} {127} (\bibinfo {year} {2014})},\ \Eprint
  {https://arxiv.org/abs/1312.3675} {arXiv:1312.3675} \BibitemShut {NoStop}%
\bibitem [{\citenamefont {Abelev}\ \emph {et~al.}(2012)\citenamefont {Abelev}
  \emph {et~al.}}]{bib:ALICE:JpsiEventAct}%
  \BibitemOpen
  \bibfield  {author} {\bibinfo {author} {\bibfnamefont {B.}~\bibnamefont
  {Abelev}} \emph {et~al.} (\bibinfo {collaboration} {ALICE Collaboration}),\
  }\href {https://doi.org/10.1016/j.physletb.2012.04.052} {\bibfield  {journal}
  {\bibinfo  {journal} {Physics Letters B}\ }\textbf {\bibinfo {volume}
  {712}},\ \bibinfo {pages} {165} (\bibinfo {year} {2012})},\ \Eprint
  {https://arxiv.org/abs/1202.2816} {arXiv:1202.2816} \BibitemShut {NoStop}%
\bibitem [{\citenamefont {Acharya}\ \emph {et~al.}(2020)\citenamefont {Acharya}
  \emph {et~al.}}]{bib:Jpsi:ALICE:EvAct13TeV}%
  \BibitemOpen
  \bibfield  {author} {\bibinfo {author} {\bibfnamefont {S.}~\bibnamefont
  {Acharya}} \emph {et~al.} (\bibinfo {collaboration} {ALICE Collaboration}),\
  }\href {https://doi.org/10.1016/j.physletb.2020.135758} {\bibfield  {journal}
  {\bibinfo  {journal} {Physics Letters B}\ }\textbf {\bibinfo {volume}
  {810}},\ \bibinfo {pages} {135758} (\bibinfo {year} {2020})},\ \Eprint
  {https://arxiv.org/abs/2005.11123} {arXiv:2005.11123} \BibitemShut {NoStop}%
\bibitem [{\citenamefont {Adam}\ \emph {et~al.}(2018)\citenamefont {Adam} \emph
  {et~al.}}]{bib:Jpsi:pp:STAR:mult}%
  \BibitemOpen
  \bibfield  {author} {\bibinfo {author} {\bibfnamefont {J.}~\bibnamefont
  {Adam}} \emph {et~al.} (\bibinfo {collaboration} {STAR Collaboration}),\
  }\href {https://doi.org/10.1016/j.physletb.2018.09.029} {\bibfield  {journal}
  {\bibinfo  {journal} {Physics Letters B}\ }\textbf {\bibinfo {volume}
  {786}},\ \bibinfo {pages} {87} (\bibinfo {year} {2018})},\ \Eprint
  {https://arxiv.org/abs/1805.03745} {arXiv:1805.03745} \BibitemShut {NoStop}%
\bibitem [{\citenamefont {Chatrchyan}\ \emph {et~al.}(2014)\citenamefont
  {Chatrchyan} \emph {et~al.}}]{bib:Ups:CMSactivity}%
  \BibitemOpen
  \bibfield  {author} {\bibinfo {author} {\bibfnamefont {S.}~\bibnamefont
  {Chatrchyan}} \emph {et~al.} (\bibinfo {collaboration} {CMS Collaboration}),\
  }\href {https://doi.org/10.1007/JHEP04(2014)103} {\bibfield  {journal}
  {\bibinfo  {journal} {Journal of High Energy Physics}\ }\textbf {\bibinfo
  {volume} {2014}},\ \bibinfo {pages} {103} (\bibinfo {year} {2014})},\ \Eprint
  {https://arxiv.org/abs/1312.6300} {arXiv:1312.6300} \BibitemShut {NoStop}%
\bibitem [{\citenamefont {Ferreiro}\ and\ \citenamefont
  {Pajares}(2012)}]{bib:PercolationJpsi}%
  \BibitemOpen
  \bibfield  {author} {\bibinfo {author} {\bibfnamefont {E.~G.}\ \bibnamefont
  {Ferreiro}}\ and\ \bibinfo {author} {\bibfnamefont {C.}~\bibnamefont
  {Pajares}},\ }\href {https://doi.org/10.1103/PhysRevC.86.034903} {\bibfield
  {journal} {\bibinfo  {journal} {Physical Review C}\ }\textbf {\bibinfo
  {volume} {86}},\ \bibinfo {pages} {034903} (\bibinfo {year} {2012})},\
  \Eprint {https://arxiv.org/abs/1203.5936} {arXiv:1203.5936} \BibitemShut
  {NoStop}%
\bibitem [{\citenamefont {Ma}\ \emph {et~al.}(2018)\citenamefont {Ma},
  \citenamefont {Tribedy}, \citenamefont {Venugopalan},\ and\ \citenamefont
  {Watanabe}}]{bib:EvAct:CGCwatanabe}%
  \BibitemOpen
  \bibfield  {author} {\bibinfo {author} {\bibfnamefont {Y.-Q.}\ \bibnamefont
  {Ma}}, \bibinfo {author} {\bibfnamefont {P.}~\bibnamefont {Tribedy}},
  \bibinfo {author} {\bibfnamefont {R.}~\bibnamefont {Venugopalan}},\ and\
  \bibinfo {author} {\bibfnamefont {K.}~\bibnamefont {Watanabe}},\ }\href
  {https://doi.org/10.1016/j.nuclphysa.2018.10.006} {\bibfield  {journal}
  {\bibinfo  {journal} {Nuclear Physics A}\ }\textbf {\bibinfo {volume}
  {982}},\ \bibinfo {pages} {747} (\bibinfo {year} {2018})},\ \Eprint
  {https://arxiv.org/abs/1807.05655} {arXiv:1807.05655} \BibitemShut {NoStop}%
\bibitem [{\citenamefont {Levin}\ and\ \citenamefont
  {Siddikov}(2019)}]{bib:EventAct:Jpsi:3pom}%
  \BibitemOpen
  \bibfield  {author} {\bibinfo {author} {\bibfnamefont {E.}~\bibnamefont
  {Levin}}\ and\ \bibinfo {author} {\bibfnamefont {M.}~\bibnamefont
  {Siddikov}},\ }\href {https://doi.org/10.1140/epjc/s10052-019-6894-1}
  {\bibfield  {journal} {\bibinfo  {journal} {The European Physical Journal C}\
  }\textbf {\bibinfo {volume} {79}},\ \bibinfo {pages} {376} (\bibinfo {year}
  {2019})},\ \Eprint {https://arxiv.org/abs/1812.06783} {arXiv:1812.06783}
  \BibitemShut {NoStop}%
\bibitem [{\citenamefont {Levin}\ \emph {et~al.}(2020)\citenamefont {Levin},
  \citenamefont {Schmidt},\ and\ \citenamefont
  {Siddikov}}]{bib:EventAct:QQ:CGC}%
  \BibitemOpen
  \bibfield  {author} {\bibinfo {author} {\bibfnamefont {E.}~\bibnamefont
  {Levin}}, \bibinfo {author} {\bibfnamefont {I.}~\bibnamefont {Schmidt}},\
  and\ \bibinfo {author} {\bibfnamefont {M.}~\bibnamefont {Siddikov}},\ }\href
  {https://doi.org/10.1140/epjc/s10052-020-8086-4} {\bibfield  {journal}
  {\bibinfo  {journal} {The European Physical Journal C}\ }\textbf {\bibinfo
  {volume} {80}},\ \bibinfo {pages} {560} (\bibinfo {year} {2020})},\ \Eprint
  {https://arxiv.org/abs/1910.13579} {arXiv:1910.13579} \BibitemShut {NoStop}%
\bibitem [{\citenamefont {Adam}\ \emph {et~al.}(2015)\citenamefont {Adam} \emph
  {et~al.}}]{bib:ALICE:CBactivity}%
  \BibitemOpen
  \bibfield  {author} {\bibinfo {author} {\bibfnamefont {J.}~\bibnamefont
  {Adam}} \emph {et~al.} (\bibinfo {collaboration} {ALICE Collaboration}),\
  }\href {https://doi.org/10.1007/JHEP09(2015)148} {\bibfield  {journal}
  {\bibinfo  {journal} {Journal of High Energy Physics}\ }\textbf {\bibinfo
  {volume} {2015}},\ \bibinfo {pages} {148} (\bibinfo {year} {2015})},\ \Eprint
  {https://arxiv.org/abs/1505.00664} {arXiv:1505.00664 [nucl-ex]} \BibitemShut
  {NoStop}%
\bibitem [{\citenamefont {Kopeliovich}\ \emph {et~al.}(2013)\citenamefont
  {Kopeliovich}, \citenamefont {Pirner}, \citenamefont {Potashnikova},
  \citenamefont {Reygers},\ and\ \citenamefont
  {Schmidt}}]{bib:Jpsi:EvAct:Kopeliovich:CPP}%
  \BibitemOpen
  \bibfield  {author} {\bibinfo {author} {\bibfnamefont {B.~Z.}\ \bibnamefont
  {Kopeliovich}}, \bibinfo {author} {\bibfnamefont {H.~J.}\ \bibnamefont
  {Pirner}}, \bibinfo {author} {\bibfnamefont {I.~K.}\ \bibnamefont
  {Potashnikova}}, \bibinfo {author} {\bibfnamefont {K.}~\bibnamefont
  {Reygers}},\ and\ \bibinfo {author} {\bibfnamefont {I.}~\bibnamefont
  {Schmidt}},\ }\href {https://doi.org/10.1103/PhysRevD.88.116002} {\bibfield
  {journal} {\bibinfo  {journal} {Physical Review D}\ }\textbf {\bibinfo
  {volume} {88}},\ \bibinfo {pages} {116002} (\bibinfo {year} {2013})},\
  \Eprint {https://arxiv.org/abs/1308.3638} {arXiv:1308.3638} \BibitemShut
  {NoStop}%
\bibitem [{\citenamefont {Kopeliovich}\ \emph {et~al.}(2020)\citenamefont
  {Kopeliovich}, \citenamefont {Pirner}, \citenamefont {Potashnikova},
  \citenamefont {Reygers},\ and\ \citenamefont
  {Schmidt}}]{bib:Jpsi:EvAct:Kopeliovich}%
  \BibitemOpen
  \bibfield  {author} {\bibinfo {author} {\bibfnamefont {B.~Z.}\ \bibnamefont
  {Kopeliovich}}, \bibinfo {author} {\bibfnamefont {H.~J.}\ \bibnamefont
  {Pirner}}, \bibinfo {author} {\bibfnamefont {I.~K.}\ \bibnamefont
  {Potashnikova}}, \bibinfo {author} {\bibfnamefont {K.}~\bibnamefont
  {Reygers}},\ and\ \bibinfo {author} {\bibfnamefont {I.}~\bibnamefont
  {Schmidt}},\ }\href {https://doi.org/10.1103/PhysRevD.101.054023} {\bibfield
  {journal} {\bibinfo  {journal} {Physical Review D}\ }\textbf {\bibinfo
  {volume} {101}},\ \bibinfo {pages} {054023} (\bibinfo {year} {2020})},\
  \Eprint {https://arxiv.org/abs/1910.09682} {arXiv:1910.09682} \BibitemShut
  {NoStop}%
\bibitem [{\citenamefont {Ackermann}\ \emph {et~al.}(2003)\citenamefont
  {Ackermann} \emph {et~al.}}]{bib:tech_STAR}%
  \BibitemOpen
  \bibfield  {author} {\bibinfo {author} {\bibfnamefont {K.~H.}\ \bibnamefont
  {Ackermann}} \emph {et~al.} (\bibinfo {collaboration} {STAR Collaboration}),\
  }\href {https://doi.org/10.1016/S0168-9002(02)01960-5} {\bibfield  {journal}
  {\bibinfo  {journal} {Nuclear Instruments and Methods in Physics Research -
  Section A}\ }\textbf {\bibinfo {volume} {499}},\ \bibinfo {pages} {624}
  (\bibinfo {year} {2003})}\BibitemShut {NoStop}%
\bibitem [{\citenamefont {Beddo}\ \emph {et~al.}(2003)\citenamefont {Beddo}
  \emph {et~al.}}]{bib:tech_BEMC}%
  \BibitemOpen
  \bibfield  {author} {\bibinfo {author} {\bibfnamefont {M.}~\bibnamefont
  {Beddo}} \emph {et~al.} (\bibinfo {collaboration} {STAR Collaboration}),\
  }\href {https://doi.org/10.1016/S0168-9002(02)01970-8} {\bibfield  {journal}
  {\bibinfo  {journal} {Nuclear Instruments and Methods in Physics Research -
  Section A}\ }\textbf {\bibinfo {volume} {499}},\ \bibinfo {pages} {725}
  (\bibinfo {year} {2003})}\BibitemShut {NoStop}%
\bibitem [{\citenamefont {Whitten}\ \emph {et~al.}(2008)\citenamefont
  {Whitten}, \citenamefont {Kponou}, \citenamefont {Makdisi},\ and\
  \citenamefont {Zelenski}}]{bib:tech:STAR:BBCpol}%
  \BibitemOpen
  \bibfield  {author} {\bibinfo {author} {\bibfnamefont {C.~A.}\ \bibnamefont
  {Whitten}}, \bibinfo {author} {\bibfnamefont {A.}~\bibnamefont {Kponou}},
  \bibinfo {author} {\bibfnamefont {Y.}~\bibnamefont {Makdisi}},\ and\ \bibinfo
  {author} {\bibfnamefont {A.}~\bibnamefont {Zelenski}},\ }in\ \href
  {https://doi.org/10.1063/1.2888113} {\emph {\bibinfo {booktitle} {AIP
  Conference Proceedings}}},\ Vol.\ \bibinfo {volume} {980}\ (\bibinfo
  {publisher} {AIP},\ \bibinfo {year} {2008})\ pp.\ \bibinfo {pages}
  {390--396}\BibitemShut {NoStop}%
\bibitem [{\citenamefont {Anderson}\ \emph {et~al.}(2003)\citenamefont
  {Anderson} \emph {et~al.}}]{bib:tech_STARTPC}%
  \BibitemOpen
  \bibfield  {author} {\bibinfo {author} {\bibfnamefont {M.}~\bibnamefont
  {Anderson}} \emph {et~al.} (\bibinfo {collaboration} {STAR Collaboration}),\
  }\href {https://doi.org/10.1016/S0168-9002(02)01964-2} {\bibfield  {journal}
  {\bibinfo  {journal} {Nuclear Instruments and Methods in Physics Research -
  Section A}\ }\textbf {\bibinfo {volume} {499}},\ \bibinfo {pages} {659}
  (\bibinfo {year} {2003})}\BibitemShut {NoStop}%
\bibitem [{\citenamefont {Llope}\ \emph {et~al.}(2004)\citenamefont {Llope}
  \emph {et~al.}}]{bib:tech:STAR:TOFp}%
  \BibitemOpen
  \bibfield  {author} {\bibinfo {author} {\bibfnamefont {W.}~\bibnamefont
  {Llope}} \emph {et~al.},\ }\href {https://doi.org/10.1016/j.nima.2003.11.414}
  {\bibfield  {journal} {\bibinfo  {journal} {Nuclear Instruments and Methods
  in Physics Research - Section A}\ }\textbf {\bibinfo {volume} {522}},\
  \bibinfo {pages} {252} (\bibinfo {year} {2004})},\ \Eprint
  {https://arxiv.org/abs/0308022} {arXiv:0308022 [nucl-ex]} \BibitemShut
  {NoStop}%
\bibitem [{\citenamefont {Adye}(2011)}]{bib:RooUnfoldMan}%
  \BibitemOpen
  \bibfield  {author} {\bibinfo {author} {\bibfnamefont {T.}~\bibnamefont
  {Adye}},\ }\href {https://doi.org/10.5170/CERN-2011-006.313} {\bibfield
  {journal} {\bibinfo  {journal} {Proceedings of the PHYSTAT 2011 Workshop}\ ,\
  \bibinfo {pages} {6}} (\bibinfo {year} {2011})},\ \Eprint
  {https://arxiv.org/abs/1105.1160} {arXiv:1105.1160} \BibitemShut {NoStop}%
\bibitem [{\citenamefont {Navas}\ \emph {et~al.}(2024)\citenamefont {Navas}
  \emph {et~al.}}]{PDG:2024}%
  \BibitemOpen
  \bibfield  {author} {\bibinfo {author} {\bibfnamefont {S.}~\bibnamefont
  {Navas}} \emph {et~al.} (\bibinfo {collaboration} {Particle Data Group
  Collaboration}),\ }\href {https://doi.org/10.1103/PhysRevD.110.030001}
  {\bibfield  {journal} {\bibinfo  {journal} {Physical Review D}\ }\textbf
  {\bibinfo {volume} {110}},\ \bibinfo {pages} {030001} (\bibinfo {year}
  {2024})}\BibitemShut {NoStop}%
\bibitem [{\citenamefont {Bichsel}(2001)}]{bib:STAR_Bichsel}%
  \BibitemOpen
  \bibfield  {author} {\bibinfo {author} {\bibfnamefont {H.}~\bibnamefont
  {Bichsel}},\ }\href@noop {} {} (\bibinfo {year} {2001}),\ \bibinfo {note}
  {{STAR note SN0439:}
  \url{https://drupal.star.bnl.gov/STAR/starnotes/public/sn0439}}\BibitemShut
  {NoStop}%
\bibitem [{\citenamefont {Vogt}(2014)}]{bib:vogt:private}%
  \BibitemOpen
  \bibfield  {author} {\bibinfo {author} {\bibfnamefont {R.}~\bibnamefont
  {Vogt}},\ }\href@noop {} {}\bibinfo {howpublished} {private communication}
  (\bibinfo {year} {2014})\BibitemShut {NoStop}%
\bibitem [{\citenamefont {Verkerke}(2010)}]{bib:RooFitMan}%
  \BibitemOpen
  \bibfield  {author} {\bibinfo {author} {\bibfnamefont {W.}~\bibnamefont
  {Verkerke}},\ }\href {https://doi.org/10.1051/epjconf/20100402005} {\bibfield
   {journal} {\bibinfo  {journal} {EPJ Web of Conferences}\ }\textbf {\bibinfo
  {volume} {4}},\ \bibinfo {pages} {02005} (\bibinfo {year}
  {2010})}\BibitemShut {NoStop}%
\bibitem [{\citenamefont {Sj{\"{o}}strand}\ \emph {et~al.}(2014)\citenamefont
  {Sj{\"{o}}strand}, \citenamefont {Ask}, \citenamefont {Christiansen},
  \citenamefont {Corke}, \citenamefont {Desai}, \citenamefont {Ilten},
  \citenamefont {Mrenna}, \citenamefont {Prestel}, \citenamefont {Rasmussen},\
  and\ \citenamefont {Skands}}]{bib:tools:PYTHIA8.2}%
  \BibitemOpen
  \bibfield  {author} {\bibinfo {author} {\bibfnamefont {T.}~\bibnamefont
  {Sj{\"{o}}strand}}, \bibinfo {author} {\bibfnamefont {S.}~\bibnamefont
  {Ask}}, \bibinfo {author} {\bibfnamefont {J.~R.}\ \bibnamefont
  {Christiansen}}, \bibinfo {author} {\bibfnamefont {R.}~\bibnamefont {Corke}},
  \bibinfo {author} {\bibfnamefont {N.}~\bibnamefont {Desai}}, \bibinfo
  {author} {\bibfnamefont {P.}~\bibnamefont {Ilten}}, \bibinfo {author}
  {\bibfnamefont {S.}~\bibnamefont {Mrenna}}, \bibinfo {author} {\bibfnamefont
  {S.}~\bibnamefont {Prestel}}, \bibinfo {author} {\bibfnamefont {C.~O.}\
  \bibnamefont {Rasmussen}},\ and\ \bibinfo {author} {\bibfnamefont {P.~Z.}\
  \bibnamefont {Skands}},\ }\href {https://doi.org/10.1016/j.cpc.2015.01.024}
  {\bibfield  {journal} {\bibinfo  {journal} {Computer Physics Communications}\
  }\textbf {\bibinfo {volume} {191}},\ \bibinfo {pages} {159} (\bibinfo {year}
  {2014})},\ \Eprint {https://arxiv.org/abs/1410.3012} {arXiv:1410.3012}
  \BibitemShut {NoStop}%
\bibitem [{\citenamefont {Gaiser}(1982)}]{bib:CBfunction}%
  \BibitemOpen
  \bibfield  {author} {\bibinfo {author} {\bibfnamefont {J.}~\bibnamefont
  {Gaiser}},\ }\href@noop {} {} (\bibinfo {year} {1982}),\ \bibinfo {note}
  {{PhD thesis, Stanford University, 1982. SLAC Report SLAC-R-255}
  \url{http://www.slac.stanford.edu/cgi-wrap/getdoc/slac-r-255.pdf}}\BibitemShut
  {NoStop}%
\bibitem [{\citenamefont {Zha}\ \emph {et~al.}(2013)\citenamefont {Zha},
  \citenamefont {Yang}, \citenamefont {Huang}, \citenamefont {Ruan},
  \citenamefont {Yang}, \citenamefont {Tang},\ and\ \citenamefont
  {Xu}}]{bib:Ups:Ratios}%
  \BibitemOpen
  \bibfield  {author} {\bibinfo {author} {\bibfnamefont {W.}~\bibnamefont
  {Zha}}, \bibinfo {author} {\bibfnamefont {C.}~\bibnamefont {Yang}}, \bibinfo
  {author} {\bibfnamefont {B.}~\bibnamefont {Huang}}, \bibinfo {author}
  {\bibfnamefont {L.}~\bibnamefont {Ruan}}, \bibinfo {author} {\bibfnamefont
  {S.}~\bibnamefont {Yang}}, \bibinfo {author} {\bibfnamefont {Z.}~\bibnamefont
  {Tang}},\ and\ \bibinfo {author} {\bibfnamefont {Z.}~\bibnamefont {Xu}},\
  }\href {https://doi.org/10.1103/PhysRevC.88.067901} {\bibfield  {journal}
  {\bibinfo  {journal} {Physical Review C}\ }\textbf {\bibinfo {volume} {88}},\
  \bibinfo {pages} {067901} (\bibinfo {year} {2013})},\ \Eprint
  {https://arxiv.org/abs/1308.4720} {arXiv:1308.4720} \BibitemShut {NoStop}%
\bibitem [{\citenamefont {Kuhr}(2010)}]{bib:Ups:CDF:pol}%
  \BibitemOpen
  \bibfield  {author} {\bibinfo {author} {\bibfnamefont {T.}~\bibnamefont
  {Kuhr}},\ }\href@noop {} {\bibinfo {title} {{Upsilon polarization measurement
  at CDF}}} (\bibinfo {year} {2010}),\ \Eprint
  {https://arxiv.org/abs/1011.0161} {arXiv:1011.0161 [hep-ex]} \BibitemShut
  {NoStop}%
\bibitem [{\citenamefont {Group}\ \emph {et~al.}(2020)\citenamefont {Group},
  \citenamefont {Zyla} \emph {et~al.}}]{bib:PDG}%
  \BibitemOpen
  \bibfield  {author} {\bibinfo {author} {\bibfnamefont {P.~D.}\ \bibnamefont
  {Group}}, \bibinfo {author} {\bibfnamefont {P.~A.}\ \bibnamefont {Zyla}},
  \emph {et~al.},\ }\href {https://doi.org/10.1093/ptep/ptaa104} {\bibfield
  {journal} {\bibinfo  {journal} {Progress of Theoretical and Experimental
  Physics}\ }\textbf {\bibinfo {volume} {2020}},\ \bibinfo {pages} {083C01}
  (\bibinfo {year} {2020})}\BibitemShut {NoStop}%
\bibitem [{\citenamefont {Adam}\ \emph {et~al.}(2019)\citenamefont {Adam} \emph
  {et~al.}}]{bib:STAR:TrkEffSyst}%
  \BibitemOpen
  \bibfield  {author} {\bibinfo {author} {\bibfnamefont {J.}~\bibnamefont
  {Adam}} \emph {et~al.} (\bibinfo {collaboration} {STAR Collaboration}),\
  }\href {https://doi.org/10.1103/PhysRevD.100.052005} {\bibfield  {journal}
  {\bibinfo  {journal} {Physical Review D}\ }\textbf {\bibinfo {volume}
  {100}},\ \bibinfo {pages} {052005} (\bibinfo {year} {2019})}\BibitemShut
  {NoStop}%
\bibitem [{\citenamefont {Adamczyk}\ \emph {et~al.}(2012)\citenamefont
  {Adamczyk} \emph {et~al.}}]{bib:STAR:Drun09}%
  \BibitemOpen
  \bibfield  {author} {\bibinfo {author} {\bibfnamefont {L.}~\bibnamefont
  {Adamczyk}} \emph {et~al.} (\bibinfo {collaboration} {STAR Collaboration}),\
  }\href {https://doi.org/10.1103/PhysRevD.86.072013} {\bibfield  {journal}
  {\bibinfo  {journal} {Physical Review D}\ }\textbf {\bibinfo {volume} {86}},\
  \bibinfo {pages} {072013} (\bibinfo {year} {2012})},\ \Eprint
  {https://arxiv.org/abs/1204.4244} {arXiv:1204.4244} \BibitemShut {NoStop}%
\bibitem [{\citenamefont {Adams}\ \emph {et~al.}(2003)\citenamefont {Adams}
  \emph {et~al.}}]{bib:STAR:highptRaa}%
  \BibitemOpen
  \bibfield  {author} {\bibinfo {author} {\bibfnamefont {J.}~\bibnamefont
  {Adams}} \emph {et~al.} (\bibinfo {collaboration} {STAR Collaboration}),\
  }\href {https://doi.org/10.1103/PhysRevLett.91.172302} {\bibfield  {journal}
  {\bibinfo  {journal} {Physical Review Letters}\ }\textbf {\bibinfo {volume}
  {91}},\ \bibinfo {pages} {172302} (\bibinfo {year} {2003})},\ \Eprint
  {https://arxiv.org/abs/0305015} {arXiv:0305015 [nucl-ex]} \BibitemShut
  {NoStop}%
\bibitem [{\citenamefont {Abelev}\ \emph {et~al.}(2006)\citenamefont {Abelev}
  \emph {et~al.}}]{bib:STAR:Long2Spin}%
  \BibitemOpen
  \bibfield  {author} {\bibinfo {author} {\bibfnamefont {B.~I.}\ \bibnamefont
  {Abelev}} \emph {et~al.} (\bibinfo {collaboration} {STAR Collaboration}),\
  }\href {https://doi.org/10.1103/PhysRevLett.97.252001} {\bibfield  {journal}
  {\bibinfo  {journal} {Physical Review Letters}\ }\textbf {\bibinfo {volume}
  {97}},\ \bibinfo {pages} {252001} (\bibinfo {year} {2006})},\ \Eprint
  {https://arxiv.org/abs/0608030} {arXiv:0608030 [hep-ex]} \BibitemShut
  {NoStop}%
\bibitem [{\citenamefont {Corke}\ and\ \citenamefont
  {Sjöstrand}(2011)}]{bib:pythia:4Cx}%
  \BibitemOpen
  \bibfield  {author} {\bibinfo {author} {\bibfnamefont {R.}~\bibnamefont
  {Corke}}\ and\ \bibinfo {author} {\bibfnamefont {T.}~\bibnamefont
  {Sjöstrand}},\ }\href {https://doi.org/10.1007/jhep05(2011)009} {\bibfield
  {journal} {\bibinfo  {journal} {Journal of High Energy Physics}\ }\textbf
  {\bibinfo {volume} {2011}},\ \bibinfo {pages} {9} (\bibinfo {year}
  {2011})}\BibitemShut {NoStop}%
\bibitem [{\citenamefont {Childress}\ \emph {et~al.}(1985)\citenamefont
  {Childress} \emph {et~al.}}]{bib:Ups:CFSpFe}%
  \BibitemOpen
  \bibfield  {author} {\bibinfo {author} {\bibfnamefont {S.}~\bibnamefont
  {Childress}} \emph {et~al.},\ }\href
  {https://doi.org/10.1103/PhysRevLett.55.1962} {\bibfield  {journal} {\bibinfo
   {journal} {Physical Review Letters}\ }\textbf {\bibinfo {volume} {55}},\
  \bibinfo {pages} {1962} (\bibinfo {year} {1985})}\BibitemShut {NoStop}%
\bibitem [{\citenamefont {Ueno}\ \emph {et~al.}(1979)\citenamefont {Ueno} \emph
  {et~al.}}]{bib:Ups:CFSppt}%
  \BibitemOpen
  \bibfield  {author} {\bibinfo {author} {\bibfnamefont {K.}~\bibnamefont
  {Ueno}} \emph {et~al.},\ }\href {https://doi.org/10.1103/PhysRevLett.42.486}
  {\bibfield  {journal} {\bibinfo  {journal} {Physical Review Letters}\
  }\textbf {\bibinfo {volume} {42}},\ \bibinfo {pages} {486} (\bibinfo {year}
  {1979})}\BibitemShut {NoStop}%
\bibitem [{\citenamefont {Innes}\ \emph {et~al.}(1977)\citenamefont {Innes}
  \emph {et~al.}}]{bib:Ups:CFSpPtCu}%
  \BibitemOpen
  \bibfield  {author} {\bibinfo {author} {\bibfnamefont {W.~R.}\ \bibnamefont
  {Innes}} \emph {et~al.},\ }\href
  {https://doi.org/10.1103/PhysRevLett.39.1240} {\bibfield  {journal} {\bibinfo
   {journal} {Physical Review Letters}\ }\textbf {\bibinfo {volume} {39}},\
  \bibinfo {pages} {1240} (\bibinfo {year} {1977})}\BibitemShut {NoStop}%
\bibitem [{\citenamefont {Moreno}\ \emph {et~al.}(1991)\citenamefont {Moreno}
  \emph {et~al.}}]{bib:Ups:E605_pCu}%
  \BibitemOpen
  \bibfield  {author} {\bibinfo {author} {\bibfnamefont {G.}~\bibnamefont
  {Moreno}} \emph {et~al.},\ }\href {https://doi.org/10.1103/PhysRevD.43.2815}
  {\bibfield  {journal} {\bibinfo  {journal} {Physical Review D}\ }\textbf
  {\bibinfo {volume} {43}},\ \bibinfo {pages} {2815} (\bibinfo {year}
  {1991})}\BibitemShut {NoStop}%
\bibitem [{\citenamefont {Yoshida}\ \emph {et~al.}(1989)\citenamefont {Yoshida}
  \emph {et~al.}}]{bib:Ups:E605_pBe}%
  \BibitemOpen
  \bibfield  {author} {\bibinfo {author} {\bibfnamefont {T.}~\bibnamefont
  {Yoshida}} \emph {et~al.} (\bibinfo {collaboration} {(E605 Collab.)}),\
  }\href@noop {} {\bibfield  {journal} {\bibinfo  {journal} {Phys. Rev. D}\
  }\textbf {\bibinfo {volume} {39}},\ \bibinfo {pages} {3516} (\bibinfo {year}
  {1989})}\BibitemShut {NoStop}%
\bibitem [{\citenamefont {Zhu}\ \emph {et~al.}(2008)\citenamefont {Zhu} \emph
  {et~al.}}]{bib:Ups:E866}%
  \BibitemOpen
  \bibfield  {author} {\bibinfo {author} {\bibfnamefont {L.~Y.}\ \bibnamefont
  {Zhu}} \emph {et~al.} (\bibinfo {collaboration} {FNAL E866/NuSea
  Collaboration}),\ }\href {https://doi.org/10.1103/PhysRevLett.100.062301}
  {\bibfield  {journal} {\bibinfo  {journal} {Physical Review Letters}\
  }\textbf {\bibinfo {volume} {100}},\ \bibinfo {pages} {062301} (\bibinfo
  {year} {2008})},\ \Eprint {https://arxiv.org/abs/0710.2344} {arXiv:0710.2344}
  \BibitemShut {NoStop}%
\bibitem [{\citenamefont {Frawley}\ \emph {et~al.}(2008)\citenamefont
  {Frawley}, \citenamefont {Ullrich},\ and\ \citenamefont
  {Vogt}}]{bib:Frawley2008}%
  \BibitemOpen
  \bibfield  {author} {\bibinfo {author} {\bibfnamefont {A.~D.}\ \bibnamefont
  {Frawley}}, \bibinfo {author} {\bibfnamefont {T.}~\bibnamefont {Ullrich}},\
  and\ \bibinfo {author} {\bibfnamefont {R.}~\bibnamefont {Vogt}},\ }\href
  {https://doi.org/10.1016/j.physrep.2008.04.002} {\bibfield  {journal}
  {\bibinfo  {journal} {Physics Reports}\ }\textbf {\bibinfo {volume} {462}},\
  \bibinfo {pages} {125} (\bibinfo {year} {2008})},\ \Eprint
  {https://arxiv.org/abs/0806.1013} {arXiv:0806.1013} \BibitemShut {NoStop}%
\bibitem [{\citenamefont {Feng}\ \emph {et~al.}(2015)\citenamefont {Feng},
  \citenamefont {Lansberg},\ and\ \citenamefont {Wang}}]{bib:lansberg:energy}%
  \BibitemOpen
  \bibfield  {author} {\bibinfo {author} {\bibfnamefont {Y.}~\bibnamefont
  {Feng}}, \bibinfo {author} {\bibfnamefont {J.-P.}\ \bibnamefont {Lansberg}},\
  and\ \bibinfo {author} {\bibfnamefont {J.-X.}\ \bibnamefont {Wang}},\ }\href
  {https://doi.org/10.1140/epjc/s10052-015-3527-1} {\bibfield  {journal}
  {\bibinfo  {journal} {The European Physical Journal C}\ }\textbf {\bibinfo
  {volume} {75}},\ \bibinfo {pages} {313} (\bibinfo {year} {2015})},\ \Eprint
  {https://arxiv.org/abs/1504.00317} {arXiv:1504.00317} \BibitemShut {NoStop}%
\bibitem [{\citenamefont {Lansberg}(2022)}]{bib:lansberg:private}%
  \BibitemOpen
  \bibfield  {author} {\bibinfo {author} {\bibfnamefont {J.-P.}\ \bibnamefont
  {Lansberg}},\ }\href@noop {} {}\bibinfo {howpublished} {private
  communication} (\bibinfo {year} {2022})\BibitemShut {NoStop}%
\bibitem [{\citenamefont {Lansberg}\ and\ \citenamefont
  {Ozcelik}(2021)}]{bib:lansberg:curingNLO}%
  \BibitemOpen
  \bibfield  {author} {\bibinfo {author} {\bibfnamefont {J.-P.}\ \bibnamefont
  {Lansberg}}\ and\ \bibinfo {author} {\bibfnamefont {M.~A.}\ \bibnamefont
  {Ozcelik}},\ }\href {https://doi.org/10.1140/epjc/s10052-021-09258-7}
  {\bibfield  {journal} {\bibinfo  {journal} {The European Physical Journal C}\
  }\textbf {\bibinfo {volume} {81}},\ \bibinfo {pages} {497} (\bibinfo {year}
  {2021})}\BibitemShut {NoStop}%
\bibitem [{\citenamefont {Lansberg}\ \emph {et~al.}(2022)\citenamefont
  {Lansberg}, \citenamefont {Nefedov},\ and\ \citenamefont
  {Ozcelik}}]{bib:lansberg:matching}%
  \BibitemOpen
  \bibfield  {author} {\bibinfo {author} {\bibfnamefont {J.-P.}\ \bibnamefont
  {Lansberg}}, \bibinfo {author} {\bibfnamefont {M.}~\bibnamefont {Nefedov}},\
  and\ \bibinfo {author} {\bibfnamefont {M.~A.}\ \bibnamefont {Ozcelik}},\
  }\href {https://doi.org/10.1007/jhep05(2022)083} {\bibfield  {journal}
  {\bibinfo  {journal} {Journal of High Energy Physics}\ }\textbf {\bibinfo
  {volume} {2022}},\ \bibinfo {pages} {83} (\bibinfo {year} {2022})},\ \Eprint
  {https://arxiv.org/abs/2112.06789} {arXiv:2112.06789 [hep-ph]} \BibitemShut
  {NoStop}%
\bibitem [{\citenamefont {Vogt}(2015)}]{bib:CEM_shadow}%
  \BibitemOpen
  \bibfield  {author} {\bibinfo {author} {\bibfnamefont {R.}~\bibnamefont
  {Vogt}},\ }\href {https://doi.org/10.1103/PhysRevD.94.114029} {\bibfield
  {journal} {\bibinfo  {journal} {Physical Review C}\ }\textbf {\bibinfo
  {volume} {92}},\ \bibinfo {pages} {034909} (\bibinfo {year} {2015})},\
  \Eprint {https://arxiv.org/abs/1507.04418} {arXiv:1507.04418} \BibitemShut
  {NoStop}%
\bibitem [{\citenamefont {Han}\ \emph {et~al.}(2016)\citenamefont {Han},
  \citenamefont {Ma}, \citenamefont {Meng}, \citenamefont {Shao}, \citenamefont
  {Zhang},\ and\ \citenamefont {Chao}}]{bib:upsCGC}%
  \BibitemOpen
  \bibfield  {author} {\bibinfo {author} {\bibfnamefont {H.}~\bibnamefont
  {Han}}, \bibinfo {author} {\bibfnamefont {Y.-Q.}\ \bibnamefont {Ma}},
  \bibinfo {author} {\bibfnamefont {C.}~\bibnamefont {Meng}}, \bibinfo {author}
  {\bibfnamefont {H.-S.}\ \bibnamefont {Shao}}, \bibinfo {author}
  {\bibfnamefont {Y.-J.}\ \bibnamefont {Zhang}},\ and\ \bibinfo {author}
  {\bibfnamefont {K.-T.}\ \bibnamefont {Chao}},\ }\href
  {https://doi.org/10.1103/PhysRevD.94.014028} {\bibfield  {journal} {\bibinfo
  {journal} {Physical Review D}\ }\textbf {\bibinfo {volume} {94}},\ \bibinfo
  {pages} {014028} (\bibinfo {year} {2016})},\ \Eprint
  {https://arxiv.org/abs/1410.8537} {arXiv:1410.8537} \BibitemShut {NoStop}%
\bibitem [{\citenamefont {Ma}\ and\ \citenamefont
  {Venugopalan}(2014)}]{bib:jpsi_cgc}%
  \BibitemOpen
  \bibfield  {author} {\bibinfo {author} {\bibfnamefont {Y.-Q.}\ \bibnamefont
  {Ma}}\ and\ \bibinfo {author} {\bibfnamefont {R.}~\bibnamefont
  {Venugopalan}},\ }\href {https://doi.org/10.1103/PhysRevLett.113.192301}
  {\bibfield  {journal} {\bibinfo  {journal} {Physical Review Letters}\
  }\textbf {\bibinfo {volume} {113}},\ \bibinfo {pages} {192301} (\bibinfo
  {year} {2014})},\ \Eprint {https://arxiv.org/abs/1408.4075} {arXiv:1408.4075}
  \BibitemShut {NoStop}%
\bibitem [{\citenamefont {Ma}(2014)}]{bib:YQMa}%
  \BibitemOpen
  \bibfield  {author} {\bibinfo {author} {\bibfnamefont {Y.-Q.}\ \bibnamefont
  {Ma}},\ }\href@noop {} {}\bibinfo {howpublished} {private communication}
  (\bibinfo {year} {2014})\BibitemShut {NoStop}%
\bibitem [{\citenamefont {Brodsky}\ and\ \citenamefont
  {Lansberg}(2010)}]{bib:ups_csm}%
  \BibitemOpen
  \bibfield  {author} {\bibinfo {author} {\bibfnamefont {S.~J.}\ \bibnamefont
  {Brodsky}}\ and\ \bibinfo {author} {\bibfnamefont {J.-P.}\ \bibnamefont
  {Lansberg}},\ }\href {https://doi.org/10.1103/PhysRevD.81.051502} {\bibfield
  {journal} {\bibinfo  {journal} {Physical Review D}\ }\textbf {\bibinfo
  {volume} {81}},\ \bibinfo {pages} {051502} (\bibinfo {year} {2010})},\
  \Eprint {https://arxiv.org/abs/0908.0754} {arXiv:0908.0754} \BibitemShut
  {NoStop}%
\bibitem [{\citenamefont {Venugopalan}(2017)}]{bib:venugopalan:private}%
  \BibitemOpen
  \bibfield  {author} {\bibinfo {author} {\bibfnamefont {R.}~\bibnamefont
  {Venugopalan}},\ }\href@noop {} {}\bibinfo {howpublished} {private
  communication} (\bibinfo {year} {2017})\BibitemShut {NoStop}%
\bibitem [{\citenamefont {Andronic}\ \emph {et~al.}(2016)\citenamefont
  {Andronic} \emph {et~al.}}]{bib:Upsilon:Overview:Andronic}%
  \BibitemOpen
  \bibfield  {author} {\bibinfo {author} {\bibfnamefont {A.}~\bibnamefont
  {Andronic}} \emph {et~al.},\ }\href
  {https://doi.org/10.1140/epjc/s10052-015-3819-5} {\bibfield  {journal}
  {\bibinfo  {journal} {The European Physical Journal C}\ }\textbf {\bibinfo
  {volume} {76}},\ \bibinfo {pages} {107} (\bibinfo {year} {2016})},\ \Eprint
  {https://arxiv.org/abs/1506.03981} {arXiv:1506.03981} \BibitemShut {NoStop}%
\bibitem [{\citenamefont {Arleo}\ \emph {et~al.}(2010)\citenamefont {Arleo},
  \citenamefont {D'Enterria},\ and\ \citenamefont {Yoon}}]{bib:pQCD:hipt}%
  \BibitemOpen
  \bibfield  {author} {\bibinfo {author} {\bibfnamefont {F.}~\bibnamefont
  {Arleo}}, \bibinfo {author} {\bibfnamefont {D.}~\bibnamefont {D'Enterria}},\
  and\ \bibinfo {author} {\bibfnamefont {A.~S.}\ \bibnamefont {Yoon}},\ }\href
  {https://doi.org/10.1007/JHEP06(2010)035} {\bibfield  {journal} {\bibinfo
  {journal} {Journal of High Energy Physics}\ }\textbf {\bibinfo {volume}
  {2010}},\ \bibinfo {pages} {35} (\bibinfo {year} {2010})},\ \Eprint
  {https://arxiv.org/abs/1003.2963} {arXiv:1003.2963} \BibitemShut {NoStop}%
\bibitem [{\citenamefont {Berman}\ \emph {et~al.}(1971)\citenamefont {Berman},
  \citenamefont {Bjorken},\ and\ \citenamefont {Kogut}}]{bib:pQCD:Kogut}%
  \BibitemOpen
  \bibfield  {author} {\bibinfo {author} {\bibfnamefont {S.~M.}\ \bibnamefont
  {Berman}}, \bibinfo {author} {\bibfnamefont {J.~D.}\ \bibnamefont
  {Bjorken}},\ and\ \bibinfo {author} {\bibfnamefont {J.~B.}\ \bibnamefont
  {Kogut}},\ }\href {https://doi.org/10.1103/PhysRevD.4.3388} {\bibfield
  {journal} {\bibinfo  {journal} {Physical Review D}\ }\textbf {\bibinfo
  {volume} {4}},\ \bibinfo {pages} {3388} (\bibinfo {year} {1971})}\BibitemShut
  {NoStop}%
\bibitem [{\citenamefont {Adamczyk}\ \emph {et~al.}(2013)\citenamefont
  {Adamczyk} \emph {et~al.}}]{bib:Jpsi:ZeboAuAu}%
  \BibitemOpen
  \bibfield  {author} {\bibinfo {author} {\bibfnamefont {L.}~\bibnamefont
  {Adamczyk}} \emph {et~al.} (\bibinfo {collaboration} {STAR Collaboration}),\
  }\href {https://doi.org/10.1016/j.physletb.2013.04.010} {\bibfield  {journal}
  {\bibinfo  {journal} {Physics Letters B}\ }\textbf {\bibinfo {volume}
  {722}},\ \bibinfo {pages} {55} (\bibinfo {year} {2013})}\BibitemShut
  {NoStop}%
\bibitem [{\citenamefont {Lafferty}\ and\ \citenamefont
  {Wyatt}(1995)}]{bib:stat:dataPoints}%
  \BibitemOpen
  \bibfield  {author} {\bibinfo {author} {\bibfnamefont {G.}~\bibnamefont
  {Lafferty}}\ and\ \bibinfo {author} {\bibfnamefont {T.}~\bibnamefont
  {Wyatt}},\ }\href {https://doi.org/10.1016/0168-9002(94)01112-5} {\bibfield
  {journal} {\bibinfo  {journal} {Nuclear Instruments and Methods in Physics
  Research - Section A}\ }\textbf {\bibinfo {volume} {355}},\ \bibinfo {pages}
  {541} (\bibinfo {year} {1995})}\BibitemShut {NoStop}%
\bibitem [{\citenamefont {Ferreiro}\ and\ \citenamefont
  {Lansberg}(2018)}]{bib:Comover_FerreiroLansberg}%
  \BibitemOpen
  \bibfield  {author} {\bibinfo {author} {\bibfnamefont {E.~G.}\ \bibnamefont
  {Ferreiro}}\ and\ \bibinfo {author} {\bibfnamefont {J.~P.}\ \bibnamefont
  {Lansberg}},\ }\bibfield  {journal} {\bibinfo  {journal} {Journal of High
  Energy Physics}\ }\textbf {\bibinfo {volume} {2018}},\ \href
  {https://doi.org/10.1007/jhep10(2018)094} {10.1007/jhep10(2018)094} (\bibinfo
  {year} {2018})\BibitemShut {NoStop}%
\bibitem [{\citenamefont {Ferreiro}\ and\ \citenamefont
  {Lansberg}(2019)}]{bib:Comover_FerreiroLansberg_err}%
  \BibitemOpen
  \bibfield  {author} {\bibinfo {author} {\bibfnamefont {E.~G.}\ \bibnamefont
  {Ferreiro}}\ and\ \bibinfo {author} {\bibfnamefont {J.~P.}\ \bibnamefont
  {Lansberg}},\ }\bibfield  {journal} {\bibinfo  {journal} {Journal of High
  Energy Physics}\ }\textbf {\bibinfo {volume} {2019}},\ \href
  {https://doi.org/10.1007/JHEP03(2019)063} {10.1007/JHEP03(2019)063} (\bibinfo
  {year} {2019})\BibitemShut {NoStop}%
\bibitem [{\citenamefont {Ullrich}()}]{bib:STAR_HFtune}%
  \BibitemOpen
  \bibfield  {author} {\bibinfo {author} {\bibfnamefont {T.}~\bibnamefont
  {Ullrich}},\ }\href@noop {} {}\bibinfo {note}
  {\url{http://www.star.bnl.gov/protected/heavy/ullrich/pythia8/}}\BibitemShut
  {NoStop}%
\end{thebibliography}%


\end{document}